\documentclass[11pt]{article}

\usepackage[utf8]{inputenc}
\usepackage[T1]{fontenc}
\newif\ifhavelmodern
\IfFileExists{lmodern.sty}{\usepackage{lmodern}\havelmoderntrue}{\havelmodernfalse}

\usepackage{amsmath,amssymb}

\usepackage[margin=1in]{geometry}
\usepackage{graphicx}
\usepackage{caption}
\usepackage{float}
\usepackage{textcomp}
\ifhavelmodern\IfFileExists{microtype.sty}{\usepackage{microtype}}{}\fi
\usepackage{authblk}

\usepackage{longtable}
\usepackage{booktabs}
\usepackage{array}
\usepackage{calc}

\usepackage{url}
\usepackage[hidelinks,breaklinks]{hyperref}
\usepackage{xurl}

\newcommand{\hl}[1]{#1}

\DeclareUnicodeCharacter{2081}{\textsubscript{1}}
\DeclareUnicodeCharacter{2082}{\textsubscript{2}}
\DeclareUnicodeCharacter{2083}{\textsubscript{3}}
\DeclareUnicodeCharacter{2084}{\textsubscript{4}}
\DeclareUnicodeCharacter{00B9}{\textsuperscript{1}}
\DeclareUnicodeCharacter{00B2}{\textsuperscript{2}}
\DeclareUnicodeCharacter{2074}{\textsuperscript{4}}
\DeclareUnicodeCharacter{2075}{\textsuperscript{5}}
\DeclareUnicodeCharacter{207B}{\textsuperscript{-}}
\DeclareUnicodeCharacter{2212}{\ensuremath{-}}
\DeclareUnicodeCharacter{00D7}{\ensuremath{\times}}
\DeclareUnicodeCharacter{00B7}{\textperiodcentered}
\DeclareUnicodeCharacter{2022}{\textbullet}
\DeclareUnicodeCharacter{2192}{\ensuremath{\rightarrow}}
\DeclareUnicodeCharacter{2264}{\ensuremath{\leq}}
\DeclareUnicodeCharacter{2248}{\ensuremath{\approx}}
\DeclareUnicodeCharacter{0394}{\ensuremath{\Delta}}
\DeclareUnicodeCharacter{03B1}{\ensuremath{\alpha}}
\DeclareUnicodeCharacter{00B1}{\ensuremath{\pm}}
\DeclareUnicodeCharacter{00B0}{\textdegree}
\DeclareUnicodeCharacter{00C5}{\AA}
\DeclareUnicodeCharacter{2010}{-}
\DeclareUnicodeCharacter{FB00}{ff}
\DeclareUnicodeCharacter{03A3}{\ensuremath{\Sigma}}
\DeclareUnicodeCharacter{03C1}{\ensuremath{\rho}}
\DeclareUnicodeCharacter{03C3}{\ensuremath{\sigma}}
\DeclareUnicodeCharacter{2080}{\textsubscript{0}}
\DeclareUnicodeCharacter{2086}{\textsubscript{6}}
\DeclareUnicodeCharacter{2088}{\textsubscript{8}}
\DeclareUnicodeCharacter{2009}{\,}
\DeclareUnicodeCharacter{1D45F}{\ensuremath{r}}

\title{\textbf{MatCreatioNN: Machine Learning Guided Computational Discovery of Photocatalysts for Environmental Applications}}

\author[a]{Satya Kokonda\thanks{Corresponding author. E-mail: \href{mailto:kokonda.satya@charterschool.org}{kokonda.satya@charterschool.org}; alternate: \href{mailto:satyakokonda@gmail.com}{satyakokonda@gmail.com}}}
\affil[a]{Department of Chemistry, School of Engineering, Charter School of Wilmington, 100 N Dupont Rd, Wilmington, DE 19807, United States of America}

\date{}

\begin{document}
\maketitle

\begin{abstract}
\hl{The rational design of photocatalysts for environmental remediation and CO₂ conversion remains limited by the high computational cost and sparse experimental data describing multi-parameter photocatalytic behavior. This work presents an integrated machine-learning framework that couples reinforcement learning-based metal--organic framework (MOF) generation with a multi-stage Crystal Graph Convolutional Neural Network (CGCNN) prediction funnel to identify photocatalysts optimized across multiple electronic and structural features. 120,000 MOF candidates were generated and screened using 13 key descriptors, including band-gap suitability, CO₂/H₂O selectivity, adsorption energy, and structural stability. The funnel approach reduced computational cost by 4.13-fold while maintaining predictive robustness. Two top candidates, a Cr-based and a Zn-based MOF, exhibited predicted photocatalytic fitness values of} 1.70±0.25 and 1.20±0.05 \hl{fold higher respectively than benchmark materials such as PCN-224(Zr), demonstrating simultaneous improvements in light absorption, redox energetics, and framework durability. Simulated X-ray diffraction patterns confirmed strong structural agreement with experimentally synthesized MOFs, indicating high synthesizability. Post-hoc analysis revealed recurring structural motifs, such as the N262 metal cluster, that correlated strongly with high predicted photocatalytic activity. These results highlight the potential of data-driven methods to accelerate discovery of efficient and durable photocatalysts for environmental and energy-related transformations, providing a foundation for experimental realization and large-scale implementation of computationally designed MOFs.}
\end{abstract}


\begin{center}
\includegraphics[width=\linewidth]{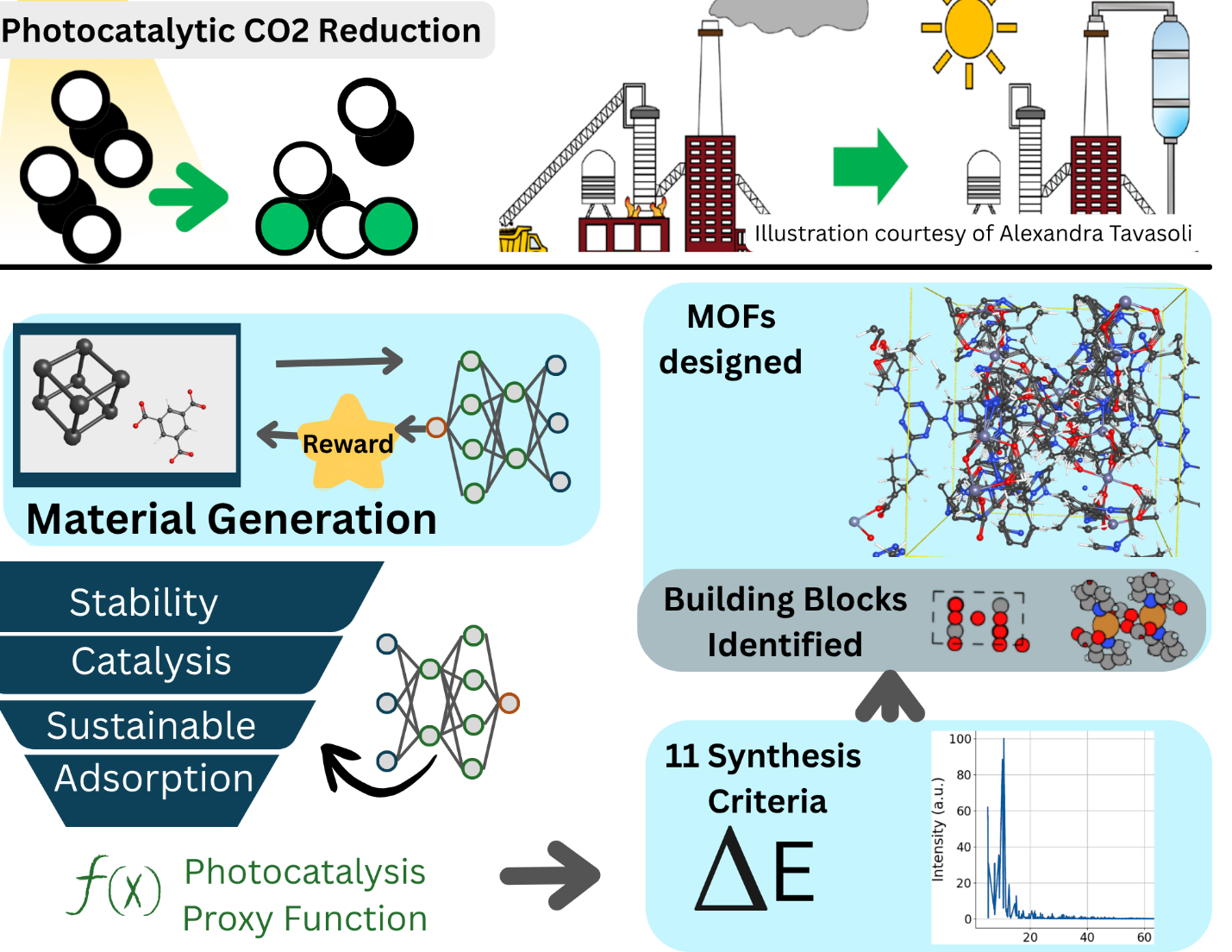}
\par\textbf{Graphical abstract.}
\end{center}

\section{Introduction}

Metal--organic frameworks (MOFs) have been recognized by the International Union of Pure and Applied Chemistry (IUPAC) as one of the ``Top Ten Emerging Technologies in Chemistry'' \cite{ref1}. These crystalline porous materials, defined by their tunable topology, metal clusters, and organic linkers, offer unparalleled chemical and structural diversity. The modular assembly of metal nodes and organic ligands enables MOFs to achieve desirable properties such as high thermal stability, adjustable porosity, and exceptionally large surface areas \cite{ref2} allowing for extensive studies in catalysis, and environmental remediation \cite{ref3}. Owing to their predictable and designable structures, MOFs permit the rational engineering of electronic, optical, and adsorption properties \cite{ref4}.

Photocatalysis is a light-driven redox process in which a semiconductor absorbs photons to generate electron--hole pairs that participate in surface oxidation and reduction reactions \cite{ref5}. The efficiency of photocatalysis is largely limited by rapid charge-carrier recombination and insufficient visible-light utilization. Constructing heterojunctions, particularly Z-scheme systems, is an effective strategy to enhance charge separation while preserving strong redox potentials \cite{ref6}.

While techniques such as Fenton chemistry and related advanced oxidation processes are highly effective for contaminant removal, they are inherently destructive and are not suitable for constructive carbon transformations such as CO₂ reduction \cite{ref7}. In contrast, photocatalytic CO₂ conversion aims to selectively activate and reduce CO₂ into value-added chemicals under light irradiation \cite{ref8}. Accordingly, oxidative remediation strategies such as Fenton chemistry are discussed here for contextual comparison but fall outside the scope of the present CO₂-reduction-focused study. Other advanced oxidation processes, including ozonation, photocatalytic oxidation, and sulfate-radical--based systems, similarly rely on highly oxidative pathways for pollutant mineralization and are therefore fundamentally distinct from the selective, reductive CO₂ conversion targeted in this work \cite{ref9}.

Photocatalysis has been applied to chemical synthesis, hydrogen production, ammonia synthesis, photoinactivation of microorganisms, and environmental remediation \cite{ref10,ref11}.

Chemical activators such as NaBH₄ and peroxymonosulfate have been reported to enhance photocatalytic degradation via radical-mediated pathways. However, these approaches rely on external reagents and reaction-specific conditions and therefore do not reflect intrinsic photocatalyst properties \cite{ref12}. Accordingly, they are not included in the present evaluation, which focuses on intrinsic, light-driven MOF performance.

Among their many applications, photocatalytic conversion of carbon dioxide (CO₂) into value-added products (e.g., CO, CH₄, or CH₃OH) offers dual benefits of greenhouse-gas mitigation and renewable-fuel generation. The photocatalytic CO₂ reduction reaction (CO₂RR) typically involves: (i) CO₂ adsorption on photocatalyst surface, (ii) photoexcitation and electron--proton transfer to form intermediates such as COOH* or CO*, and (iii) desorption of reduced product \cite{ref13}. In this context, MOF-based photocatalysts serve as light absorbers and structural hosts facilitating the diffusion and activation of reactant molecules within confined pore environments. MOFs are particularly suited for CO₂RR because their high surface area and tunable pore environment enable control over light harvesting, charge separation, and molecular confinement effects \cite{ref14}.

Conventional anaerobic digestion (AD) is widely used for waste treatment and biogas production; however, it is inherently limited for targeted CO₂ valorization and photocatalytic process development. AD operates on long residence times (days to weeks), produces low-value methane-dominated gas mixtures, and offers limited control over product selectivity and reaction pathways \cite{ref15}. In addition, AD performance is highly sensitive to feedstock variability, microbial community stability, and operating conditions, which constrains rapid optimization and materials-level innovation. In contrast, photocatalytic CO₂ conversion enables direct, light-driven activation of CO₂ with tunable selectivity toward value-added products and operates on fundamentally faster, catalyst-controlled timescales \cite{ref16}. The present work is novel in that it introduces a multi-objective, machine-learning--guided framework to rapidly prioritize photocatalyst materials for CO₂ conversion, integrating catalytic, adsorption, stability, economic, and sustainability criteria. This approach enables orders-of-magnitude acceleration of materials discovery relative to conventional trial-and-error experimentation \cite{ref17} and provides a scalable pathway to identify photocatalysts tailored for targeted CO₂ valorization rather than bulk biogas generation.

Representative MOF photocatalysts include porphyrin-based frameworks such as PCN-224 and PCN-222, which leverage metal-porphyrin chromophores for visible-light absorption and achieve CO₂-to-CO conversion with high selectivity \cite{ref18}. Zinc-based MOFs like ZIF-8 excel in stability but often have high band gap causing weak light adsorption \cite{ref19} while Chromium MOFs like MIL-101(Cr) variants offer exceptional chemical stability but limited photocatalytic activity due to rapid electron--hole recombination \cite{ref20}. However, no single framework simultaneously satisfies requirements for optimal band gap, high CO₂ selectivity, structural durability, and economic viability \cite{ref21}. The novelty of the present work lies in the use of a machine-learning--guided framework that holistically optimizes metal--organic frameworks across structural, electronic, adsorption, and stability descriptors, rather than focusing on a single oxidation pathway or isolated material property.

\section{Literature Review}

Traditional computational screening faces a combinatorial explosion: with \textgreater800 building blocks and 1,700+ topologies catalogued; the MOF design space exceeds 10\textsuperscript{14} hypothetical structures \cite{ref22}. Exhaustive DFT or Grand Canonical Monte Carlo characterization of electronic properties requires \textasciitilde100 CPU-hours per structure, rendering brute-force search infeasible \cite{ref17}. Machine learning (ML) circumvents this bottleneck by learning structure--property relationships from limited DFT training data, enabling prediction at \textless0.05 second per structure, a 10\textsuperscript{7}-fold speedup that makes high-throughput screening practical (7 minutes as compared to 1.5 million hours) \cite{ref17}.

Recent ML-driven studies have advanced MOF discovery through (i) inverse design with generative models trained on existing databases \cite{ref23} and (ii) Bayesian or genetic optimization \cite{ref24,ref25}. These, alongside most machine learning frameworks, target one or two descriptors at a time, limiting their ability to capture trade-offs between adsorption capacity, electronic structure, or stability \cite{ref26}. Additionally, these approaches are rarely validated photoactive conditions relevant to catalysis \cite{ref27}. To address these gaps, the present study couples reinforcement learning--based MOF generation (\textgreater10⁵ candidates) with a sequential Crystal Graph Convolutional Neural Network (CGCNN) funnel that simultaneously evaluates 13 photocatalytic descriptors, reducing computational cost by \textgreater4-fold while maintaining predictive accuracy.

\section{Methodology}

\begin{figure}[htbp]
\centering
\includegraphics[width=\linewidth]{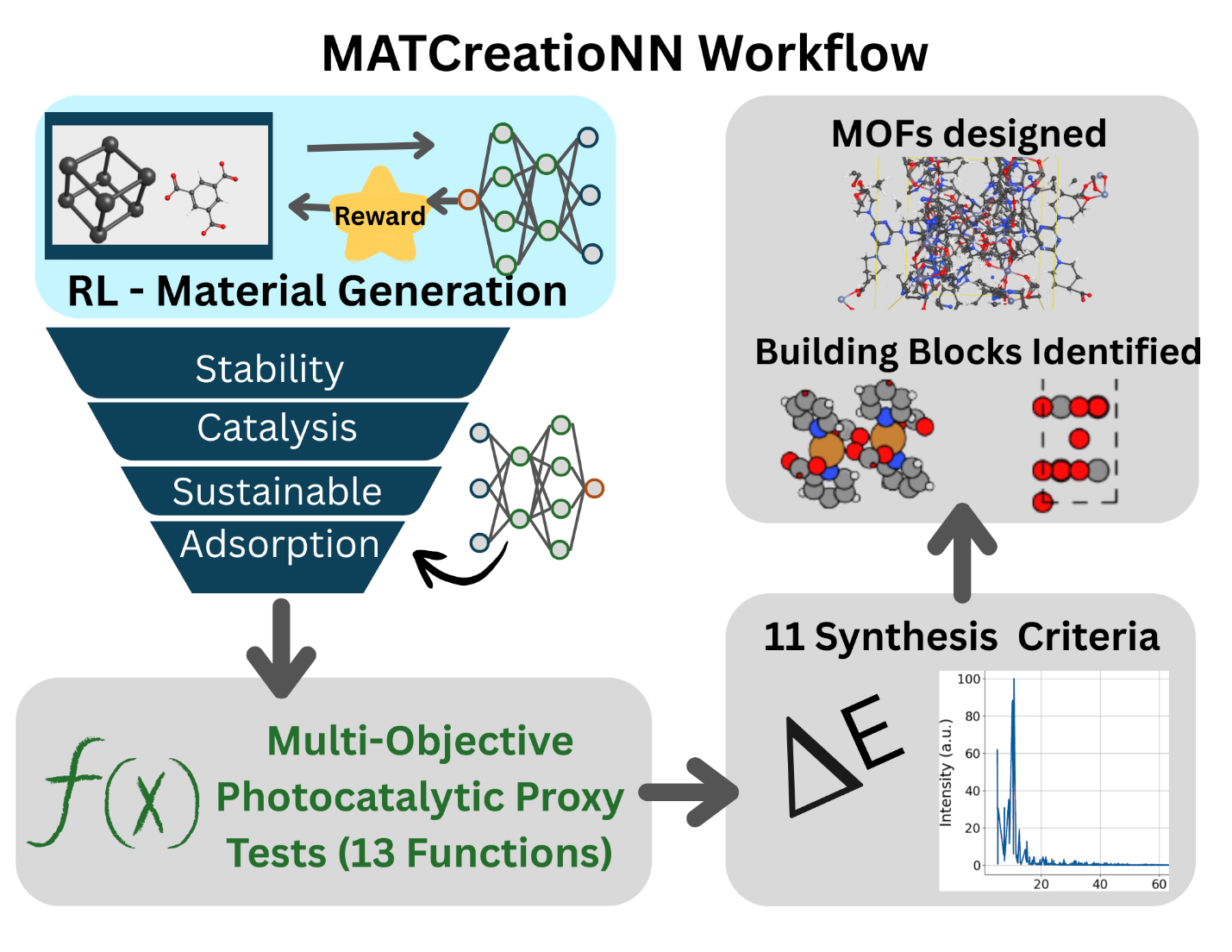}
\caption{\textbf{Overview of the MATCreatioNN machine-learning workflow for photocatalytic MOF discovery.} Reinforcement-learning--based MOF generation is followed by sequential CGCNN screening for stability, catalytic proxies, adsorption, and economic/sustainability metrics. Candidates are evaluated using an ensemble of 13 multi-objective fitness aggregation functions to ensure robustness to fitness-function choice. Top-ranked materials are further filtered by synthesis and feasibility criteria and assessed for structural plausibility.}
\label{fig:1}
\end{figure}

Full methodological details can be found in Fig.~\ref{fig:1}.

\subsection{Reinforcement Learning-Based MOF Generation}

An ensemble reinforcement learning framework \cite{ref23} generated 120,000 MOF candidates, comprising 60,000 structures optimized for CO₂ heat-of-adsorption and 60,000 for CO₂/H₂O selectivity. The encoder-decoder architecture of MOFReinforce has more detail found in Supplemental Information (S2). Various post-processing steps were used to select optimal materials \cite{ref23}. Out of those MOF candidates, 42117 failed optimizations using Universal Forcefield (UFF) \cite{ref28}. This is a widely used constraint, so generated materials align with their lowest-energy usage form. After material ranking, the lowest 15\% of materials based on their respective optimized features (Heat of Adsorption/Selectivity) were removed. The MOFs were converted into Crystallographic Information File (CIF) format, a widely adopted standard for storing molecular and crystal structure \cite{ref29}. The materials had to pass through four preliminary synthesis criteria: maintain \textless{} 3000 atoms, cell dimensions \textless{} 60 Å, synthesis accessibility score \textless{} 6, and topological root mean squared deviation (RMSD) \textless{} 0.3 Å \cite{ref23,ref30}.

\subsection{Prediction System}

Multiple features influence photocatalytic performance, but exhaustive high-throughput evaluation of all materials is computationally prohibitive. An iterative funnel system based on Crystal Graph Convolutional Neural Networks (CGCNNs) \cite{ref31} was therefore implemented to sequentially screen materials across multiple performance-relevant descriptors. Three filtering strategies were applied: (i) quantile cutoffs (5\%) for property extremes, (ii) absolute thresholds (e.g., temperature stability), and (iii) classification rules (synthesizability). This approach mitigates parameter overfitting by preventing optimization of a narrow subset of descriptors at the expense of other critical factors \cite{ref32}. A 5\% quantile cutoff was selected to balance selectivity and structural diversity (see Supplementary Information S3.1), and hyperparameters were optimized using an 80/10/10 train/validation/test split with a Pareto front minimizing benchmark MAE and inference time (Supplementary Information S3.2).

\subsubsection{Funnel Organization}

Two features were additively used to model stability. Water stability is critical because flue gas from coal-fired power plants contains 20--23\% H₂O \cite{ref33}, with stability labels derived from the MOFSimplify database \cite{ref34}. High bulk modules were included to ensure resistance to mechanical stress and structural degradation under industrial operating conditions. Together, these properties capture resistance to hydrolysis and framework collapse under humid and mildly acidic environments typical of CO₂RR and industrial AOP systems \cite{ref35}. A quantile cutoff on the combined stability score reduced the candidate pool to 33,582 MOFs.

Due to limited photocatalysis training data and mechanistic parallels in CO₂ reduction pathways (shared *COOH and *CO intermediates, metal-center redox chemistry, and similar energetic landscapes), three pre-trained electrocatalysis models (Faradaic efficiency, free energy, and voltage potential) were adapted as proxy predictors for photocatalytic activity \cite{ref36} (see Section 3.2.2). To account for photon-driven excitation, a band-gap predictor was added with optimal performance at 1.8--2.0 eV \cite{ref37}, enforced via Gaussian weighting (Supplementary Information S3.3). These four descriptors were combined additively as the catalytic score, and bottom-quantile filtering reduced the pool to 31,902 MOFs.

Two domain-specific CGCNN models were trained to predict material cost and environmental sustainability. The cost predictor was trained on 20,374 MOFs using element-based cost approximations derived from the QMOF database (Supplementary Information S3.4). The sustainability predictor, trained on 2,282 MOFs, assessed metal sourcing from industrial and post-consumer waste streams (e.g., electronic scrap, battery waste, wastewater sludge, and refinery by-products) \cite{ref38}. Sequential quantile cutoffs reduced the pool to 27,391 materials.

Post-combustion flue gas is typically cooled to approximately 148 °C (300 °F) prior to adsorption \cite{ref39}. A CGCNN thermal-stability predictor trained on the MOFSimplify database was applied with a 148 °C minimum threshold, retaining 20,943 thermally stable candidates.

Three additive metrics were used to evaluate adsorption effectiveness. The available surface area (ASA) was interpreted as a proxy for active-site density and photon utilization rather than a purely geometric property \cite{ref40}, serving as a computational surrogate for experimentally measured BET surface area. Larger and more interconnected pore networks increase the number of photoreactive sites and facilitate charge-carrier transport. The CO₂ heat of adsorption (HOA) and CO₂/H₂O selectivity were included to capture mechanistic control of CO₂ activation and competitive adsorption in humid environments. HOA values were evaluated using literature-supported windows (−20 to −40 kJ mol⁻¹, with limited viability near −60 kJ mol⁻¹) \cite{ref41} and processed using a truncated Gaussian weighting function (Supplementary Information S3.3). Combined adsorption metrics were filtered by quantile ranking, leaving 19,876 MOFs.

A CGCNN binary classifier (α = 0.5) was trained on the QMOF database \cite{ref17} to predict experimental synthesizability using 20,375 MOFs labeled as experimentally realized or hypothetical. This step retained 17,315 likely synthesizable candidates.

Prediction accuracy statistics for all models are provided in Supplementary Information S3.5. Implementation of the funnel system resulted in a 4.13-fold improvement in computational efficiency relative to brute-force high-throughput screening (Supplementary Information S3.6).

\begin{figure}[htbp]
\centering
\includegraphics[width=\linewidth]{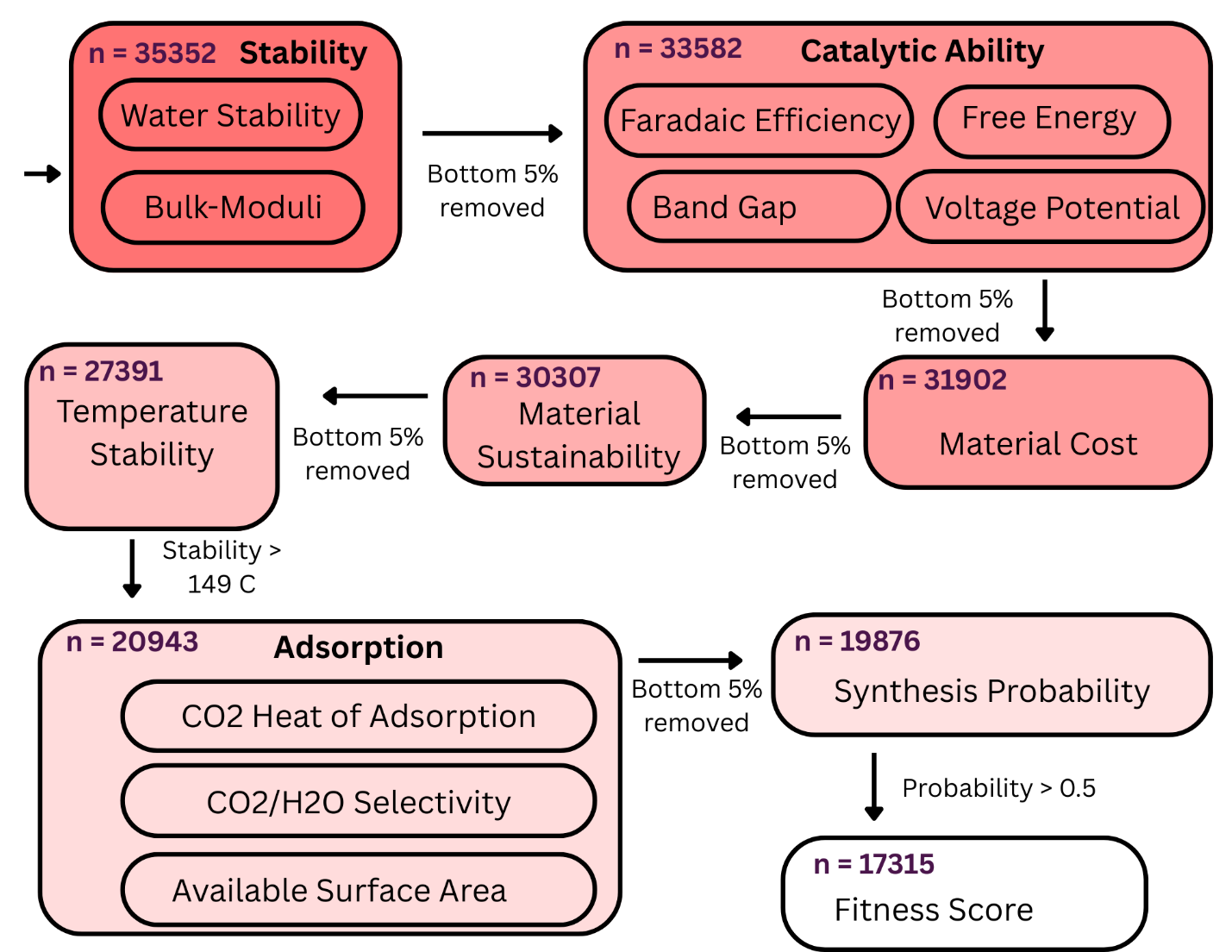}
\caption{\textbf{Workflow of the CGCNN funnel system.} Thirteen sequential CGCNN modules predict physical, electronic, and catalytic descriptors of MOFs. At each stage, the lowest 5\% of candidates are removed based on property-specific thresholds, maintaining dataset diversity while progressively enriching high-performing structures. This hierarchical filtration reduces overfitting and improves computational efficiency by 4.13-fold compared to exhaustive evaluation of all properties on all candidates (Supplemental Information for calculation details)}
\label{fig:2}
\end{figure}

\subsubsection{Justification for electrocatalysis-derived proxy descriptors}

The electrocatalysis predictors (Faradaic efficiency, free energy, and voltage potential) were incorporated as proxy descriptors for photocatalytic CO₂ reduction based on the well-established mechanistic equivalence between photo- and electro-driven CO₂ activation pathways. Although the electron source differs (photogenerated carriers versus applied potential), both catalytic modes proceed through the same sequence of elementary intermediates (*CO₂⁻, *COOH, *CO) and proton--electron transfer steps \cite{ref14,ref37}. Prior studies have shown that ΔG(*COOH) and related electrocatalytic free-energy descriptors correlate with photocatalytic CO₂RR activity in MOFs and semiconductors \cite{ref42}, and that band-edge alignment coupled with intermediate stabilization governs visible-light CO₂ conversion efficiency \cite{ref43}.

Because large, standardized photocatalysis datasets remain scarce \cite{ref44,ref45}, transfer learning from electrocatalysis has become a common strategy for CO₂RR materials discovery. Recent machine-learning studies demonstrate that electrocatalytic ΔG pathways and related descriptors successfully transfer to photocatalytic screening tasks \cite{ref46}. Consistent with this, several MOFs reported as strong photocatalysts (e.g., PCN-222, MIL-101(Cr), UiO-66 derivatives) are also strong electrocatalysts, reflecting shared structure--property relationships in metal-node chemistry, linker fields, topology, and adsorption energetics.

In the present funnel, electrocatalytic predictors are not used in isolation but are combined with band-gap weighting, CO₂/H₂O adsorption descriptors, and stability metrics to mitigate domain-mismatch bias. Ensemble fitness evaluation across thirteen independent functional forms further reduces sensitivity to any single proxy. Benchmark photocatalysts rank appropriately under this combined scoring, indicating that electrocatalytic descriptor transfer does not distort relative performance within the MOF domain.

More broadly, machine-learning frameworks commonly integrate electronic and thermodynamic descriptors across electro- and photocatalytic systems, supporting the transferability of these features \cite{ref47,ref48}. Faradaic efficiency for CO₂ reduction has also been shown to be largely independent of electron sourcing, supporting its use as a transferable proxy for photocatalytic screening \cite{ref49}.

\subsection{Material Benchmark}

To benchmark the materials generated, a control set of 8 promising MOF-based photocatalysts was compiled from recent review articles (Table S6). When multiple CIFs were available for a given structure, all corresponding files were included to reduce dataset-specific bias. Each control MOF was processed through the same CGCNN module and post-processing pipeline used for the generated set, ensuring direct comparability of fitness scores. Because machine learning models yield the most reliable predictions within their trained chemical domain, benchmarking was restricted to MOF photocatalysts to avoid extrapolation beyond the CGCNN training distribution.

Although no new computations are introduced, the curated set of literature photocatalysts ensures that the model's predictions fall within physically reasonable trends. Each control MOF is well-established in literature as a visible-light photocatalyst \cite{ref50}.

\subsection{Generalized AOP Performance Framework (GAPF)}

Multi-objective catalyst optimization faces a fundamental challenge: no single scoring function can objectively balance competing design criteria (e.g., high catalytic activity vs. low cost) \cite{ref51,ref52}. To ensure robust rankings independent of weighting assumptions, we developed an ensemble-based evaluation framework that tests MOF performance across 13 alternative fitness functions spanning different mathematical forms and parameter emphases (Supplemental Information S5.1, Equations Eq. S1- Eq. S13). Materials demonstrating consistent superiority across this diverse function space are less likely to represent artifacts of arbitrary metric choice \cite{ref53,ref54}.

Each function combines the five normalized descriptors (stability, catalytic activity, cost, sustainability, adsorption) using different mathematical forms, additive, multiplicative, harmonic mean, etc., allowing us to test whether rankings remain consistent across aggregation methods.

Full mathematical definitions of the thirteen tested aggregation models (F₁--F₁₃) are provided in Supplementary Information. Additionally, to test the effect of specific parameter weight, equation forms including catalytic emphasis, no economic terms and extreme economic form were included.

Aggregation ensemble (no single fitness). For each candidate we compute the vector

\[F(x) = (f_{1}(x),\ldots,f_{13}(x))
\]

using the canonical formulations (additive, geometric, harmonic, log-sum, min--max, etc.) defined in SI (S5.1).

Uncertainty model. Each descriptor is modeled as \(X_{i}(m_{i},{(rm_{i}\text{)}}^{\text{2}}\text{ )}\text{ }\)with \(r \in \{ 0.05,0.10,0.11,0.20\}\). For any aggregation \(f( \cdot )\), the propagated \(\sigma_{f}\)is computed using first-order Gaussian propagation \cite{ref54} with analytic \(\frac{\partial f}{\partial x_{i}}\ \). Calculations in SI (S5.1).

Uncertainty-resolved separation (Z).

\[Z_{k} = \frac{\mu\lbrack F_{k}(a)\rbrack - \mu\lbrack F_{k}(c)\rbrack}{\sqrt{\sigma^{2}\lbrack F_{k}(a)\rbrack + \sigma^{2}\lbrack F_{k}(c)\rbrack}}
\]

with \(\sigma\left\lbrack F_{k} \right\rbrack\ \)from S5.1. We summarize count of \(Z_{k} > 2\ \)across the ensemble.

A Z-score threshold of 2 was used to identify statistically significant separations (\textgreater95\% confidence) between control and modified MOFs.

An expert-informed consensus fitness function (S5.2) was constructed by consolidating the most physically meaningful descriptors identified across the 13 formulations.

This consensus function, developed under domain guidance and informed by the ensemble results, was used exclusively for post-processing to rank the 120,000 generated MOFs and identify recurring structural motifs (Section 4). It was not employed in the funnel filtering stages (Section 3.2) or in candidate-control benchmarking (Section 3.3).

\subsection{Computational X-Ray Diffraction}

Simulated XRD patterns were generated, with implementation available in the GitHub (S1). Analyses were performed for both the MOFs generated in this study and reference structures from the QMOF database \cite{ref17}. Simulated diffraction profiles were used to evaluate crystallographic consistency between newly designed and experimentally reported MOFs \cite{ref55}.

\section{Results}

Post-hoc analysis of the top 1,000 MOFs ranked by the composite fitness function revealed strong enrichment of the bcg topology and the N262 metal cluster, accounting for 34.0\% and 31.3\% of high-performing candidates, respectively. A notable drop in occurrence is found, where the most popular building-block in both scenarios is \textgreater4 fold as commonly used as the next-alternative.

\begin{figure}[htbp]
\centering
\includegraphics[width=\linewidth]{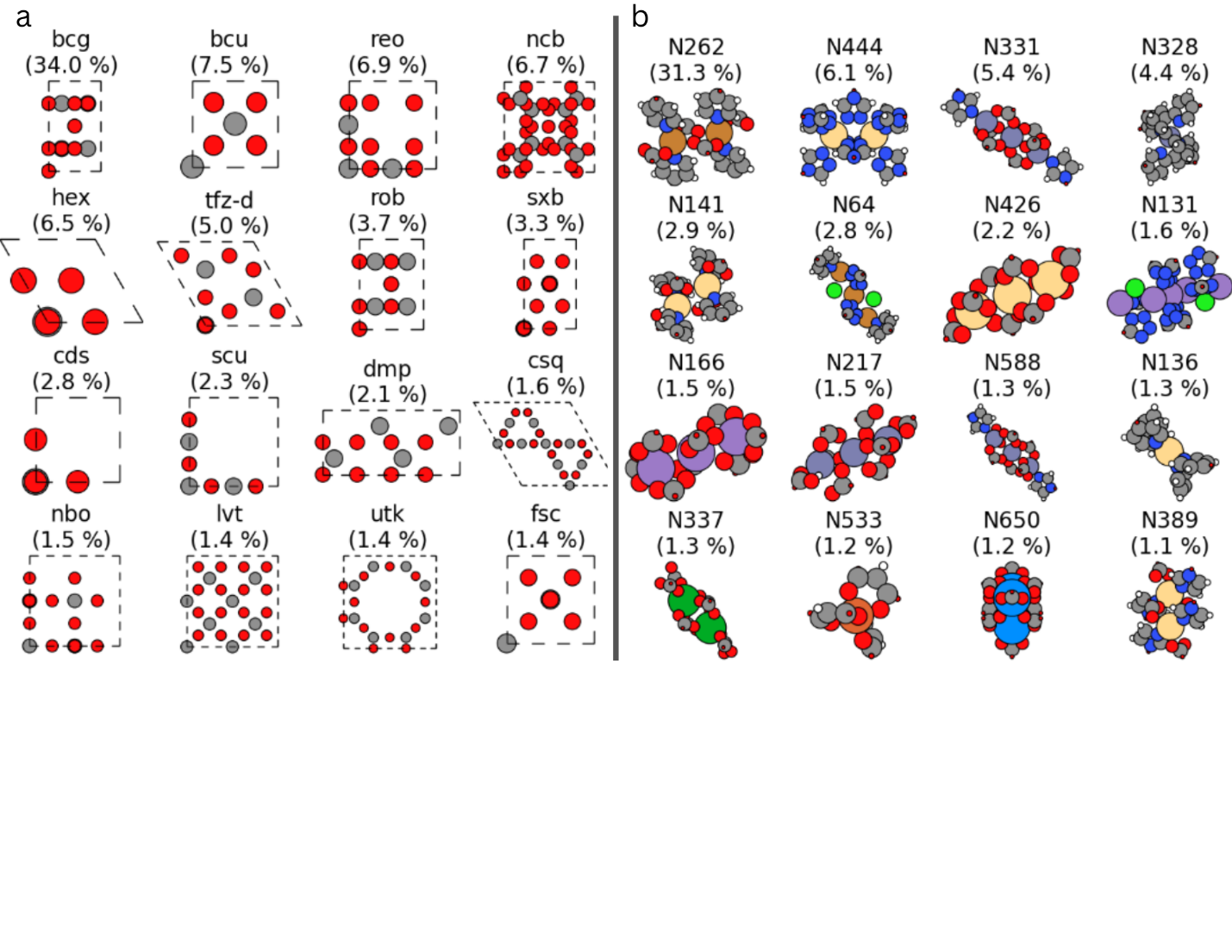}
\caption{Distribution of (a) topologies and (b) metal clusters among the 1,000 highest-ranked MOFs for photocatalytic performance, as identified by the expert-informed composite fitness function. The distinct enrichment pattern demonstrates the model's capacity to adaptively recognize recurring structural motifs, with the N262 metal cluster appearing in 31.3 \% of candidates, highlighting its significance as a potential design motif for future photocatalyst development.}
\label{fig:3}
\end{figure}

Zn-based MOF outperformed the benchmark across each individual measured parameters (Fig.~\ref{fig:4}) (Full parameter value information found in supplemental information S6.1), suggesting more robust predicted performance independent of fitness function weighting. This MOF had narrow, overlapping intervals across each estimator, indicating robust and consistent performance, while Cr displayed broader confidence ranges reflecting heterogeneity in functional gains.

\begin{figure}[htbp]
\centering
\includegraphics[width=\linewidth]{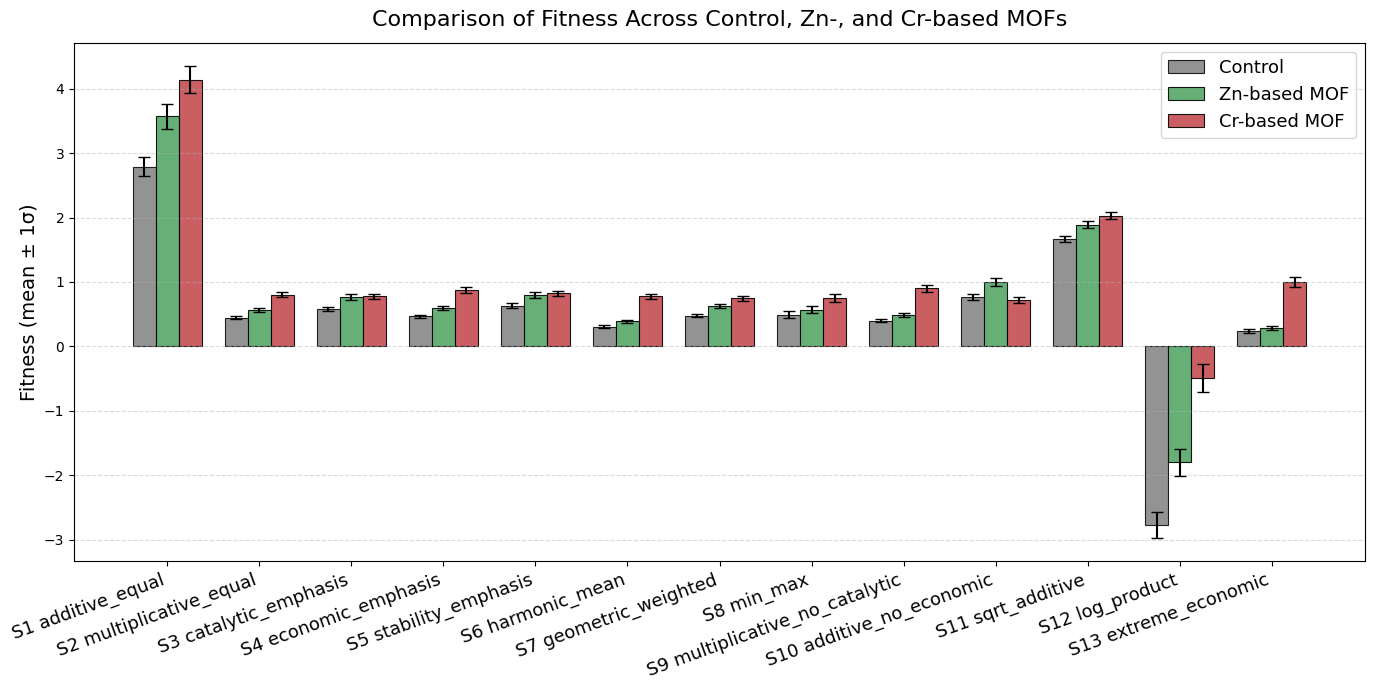}
\caption{Multi-Fitness comparison of two identified MOFs compared to general top performing control. Relative errors were propagated through each function using a r=0.11 uncertainty based on prior literature \cite{ref17}. The Zn-Based MOF outperforms the top control in measured cases, with the Cr-based MOF outperforming in 12/13. Data available in supplemental information (Table S8).}
\label{fig:4}
\end{figure}

Instead of comparing solely fitness attributes, a focus should be placed on holistic improvements to the baseline models, where we identify materials, namely the Zinc-Based MOF that surpasses the most promising control (PCN-224-Zr) in each optimized parameter measured.

\begin{figure}[htbp]
\centering
\includegraphics[width=\linewidth]{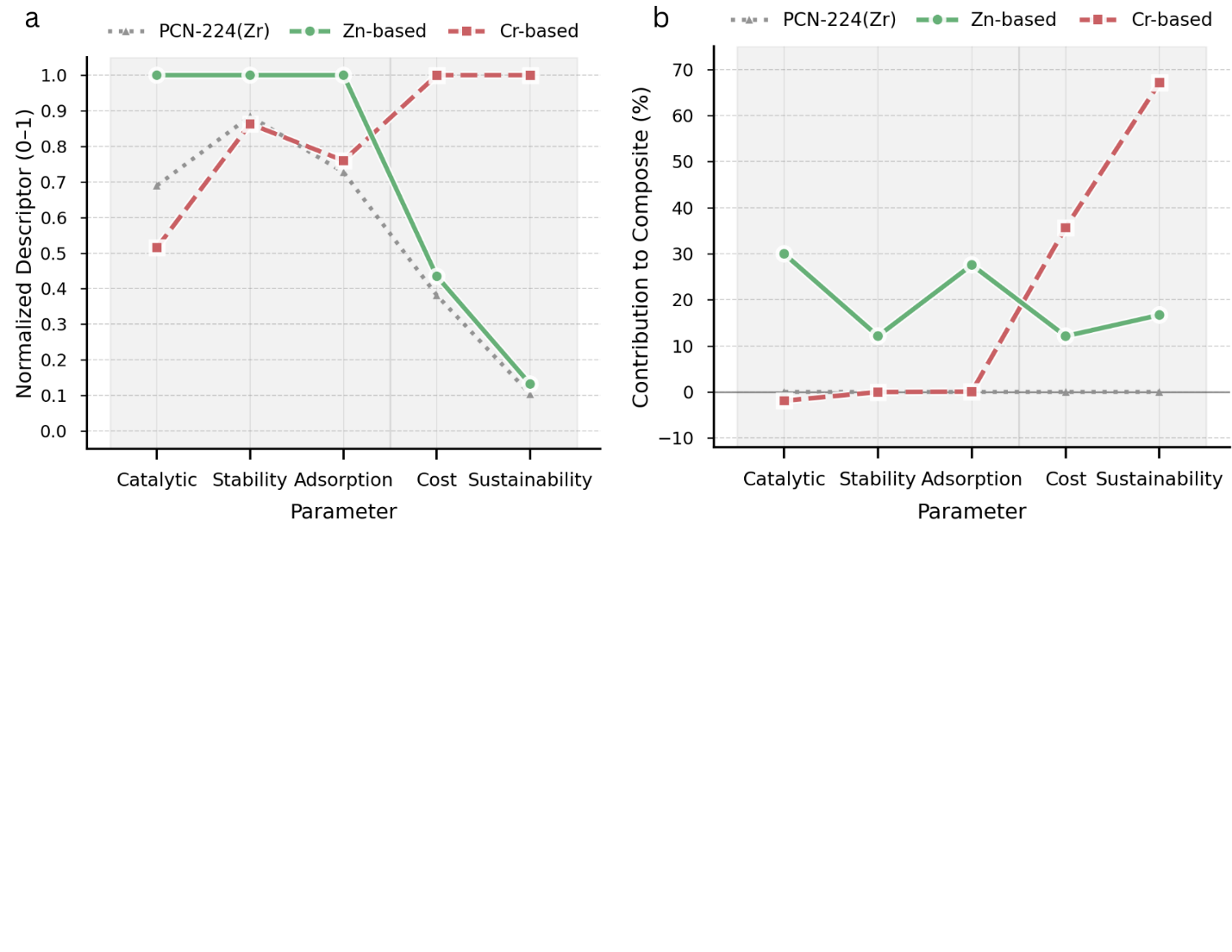}
\caption{(a) Normalized descriptor profiles (0--1 scaling) for the same MOFs, illustrating that Zn-MOF maintains high intrinsic performance metrics (Catalytic = 1.0, Adsorption = 1.0) while Cr-MOF achieves superior economic and sustainability factors (Inverse Cost = 1.0, Sustainability = 1.0) despite lower catalytic normalization. (b) Parameter-level contributions to composite fitness improvement for Zn-based and Cr-based MOFs relative to the PCN-224(Zr) benchmark. Contributions were derived from mean Shapley values, which averaged across thirteen alternative fitness aggregation models. Zn-MOF exhibits a general performance-optimized profile, whereas Cr-MOF shows an economic-optimized profile driven by Sustainability and Inverse Cost. Error bars are omitted for clarity; full uncertainty propagation and standard deviations for aggregation functions are provided in Supplementary Information.}
\label{fig:5}
\end{figure}

Bootstrap and jackknife resampling across the 13 fitness functions confirmed mean improvements of 1.20 ± 0.05× for the Zn-based MOF and 1.70 ± 0.25× for the Cr-based MOF, demonstrating statistically significant and robust performance gains relative to PCN-224(Zr) (full resampling analysis in Supplementary Information).

Rankings showed strong agreement across alternative fitness formulations, Kendall\textquotesingle s W = 0.79 indicates high robustness to aggregation method choice (see Supplementary Information for full correlation analysis).

\begin{figure}[htbp]
\centering
\includegraphics[width=\linewidth]{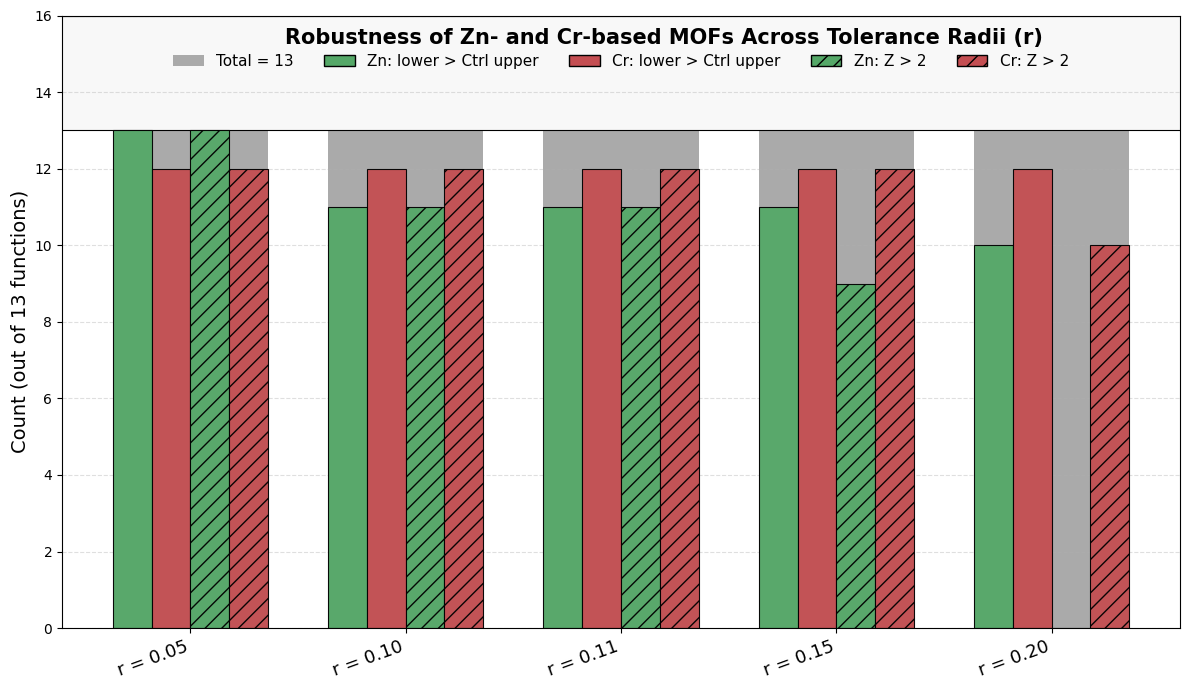}
\caption{Uncertainty propagation analysis showing robust separation of Zn- and Cr-based MOFs from the benchmark under typical CGCNN prediction errors. Full robustness analysis is provided in Supplementary Information.}
\label{fig:6}
\end{figure}

Uncertainty propagation analysis (Fig.~\ref{fig:6}) demonstrates that both top candidates remain statistically separated from the benchmark under realistic prediction error levels typical of CGCNN models (r=0.11) \cite{ref17,ref31,ref48}. Full perturbation-level analysis is provided in Supplementary Information.

Both generated MOFs exhibited very large effect sizes relative to PCN-224(Zr) across fitness functions (Cohen's d: Zn = 3.87 ± 1.82; Cr = 8.51 ± 6.34). All 13 fitness functions yielded statistically significant improvements for both Zn- and Cr-based MOFs after Bonferroni--Holm correction (corrected p \textless{} 0.001), indicating robustness to multiple-testing effects.

\begin{figure}[htbp]
\centering
\includegraphics[width=\linewidth]{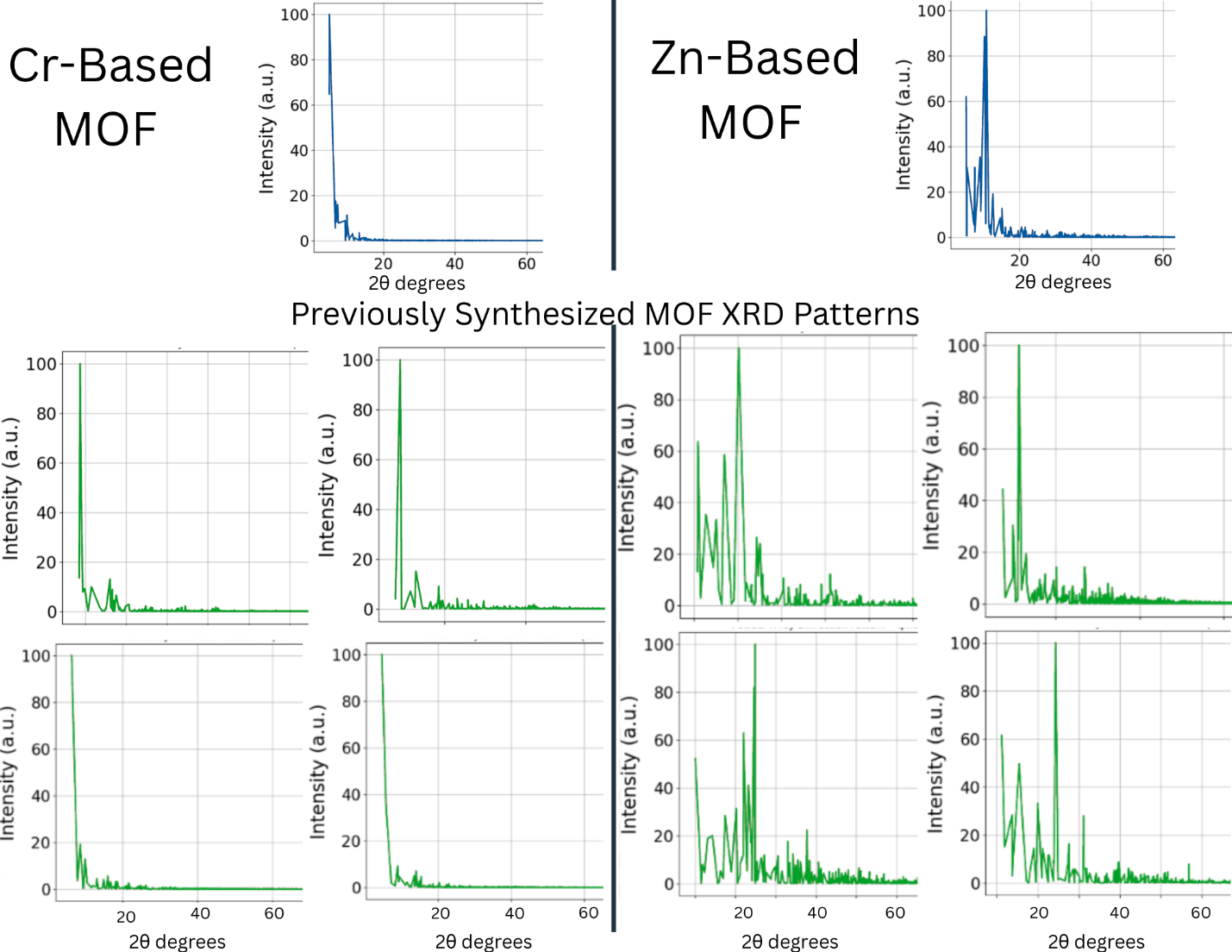}
\caption{\textbf{Simulated and experimental XRD comparison for top-performing generated MOFs.} XRD analysis was conducted on the highest-ranked predicted structures to confirm structural feasibility and compare diffraction profiles against known synthesized analogs. Close alignment between simulated and experimental peaks validates the crystallographic plausibility of the generated MOFs.}
\label{fig:7}
\end{figure}

Simulated XRD patterns (Fig.~\ref{fig:7}) confirmed crystallographic plausibility of generated structures. Both Zn-based and Cr-based MOFs exhibited diffraction profiles closely matching experimentally synthesized analogs from the QMOF database, with dominant peak positions, relative intensities, and peak counts nearly identical to known frameworks.

\begin{figure}[htbp]
\centering
\includegraphics[width=\linewidth]{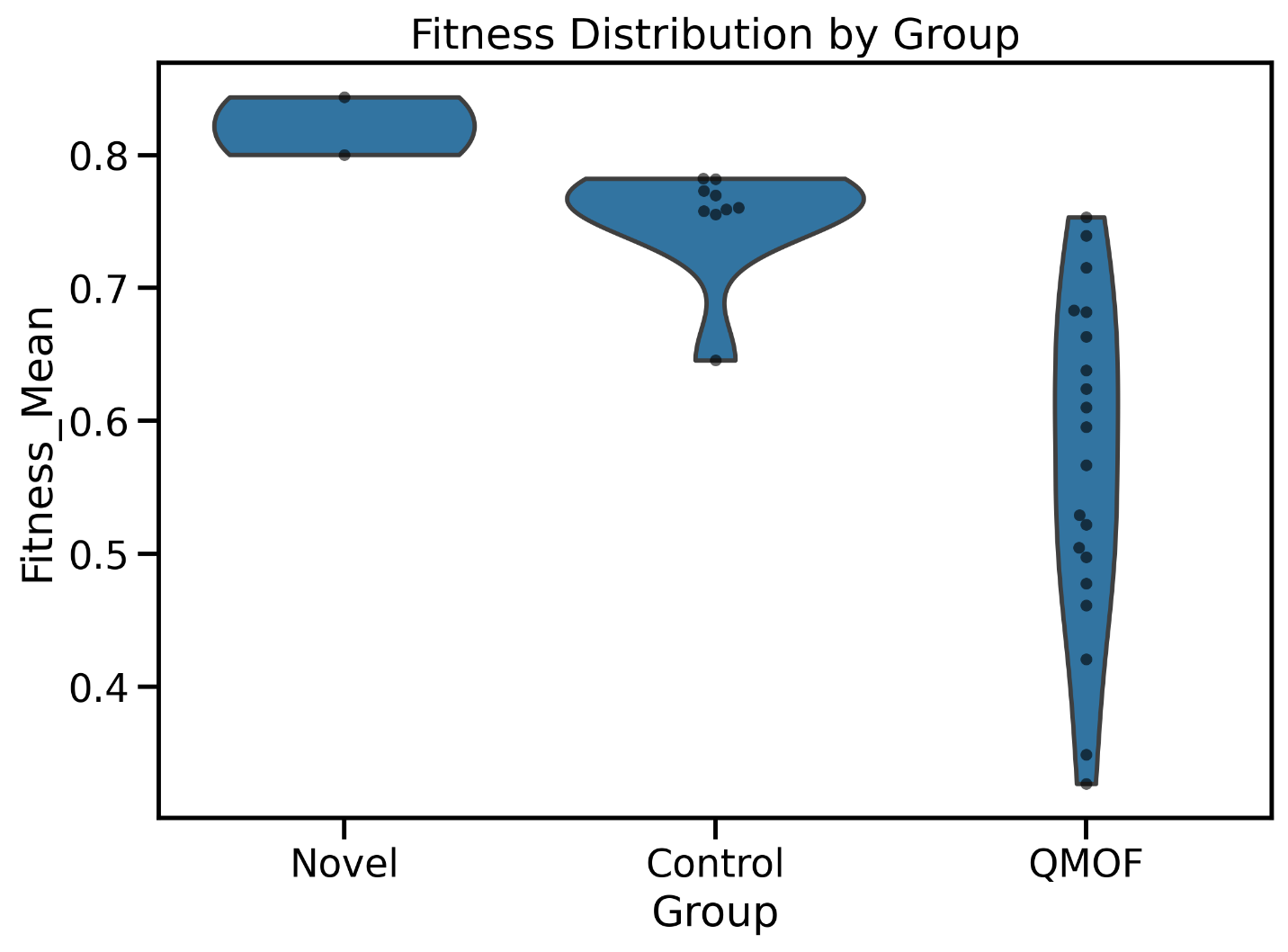}
\caption{Violin (ribbon) plots show the distribution of mean composite fitness scores (averaged over f1--f13) for generated MOFs, curated photocatalytic control MOFs, and randomly sampled QMOF structures. Individual data points are overlaid. Generated MOFs exhibit a clear right-shift relative to both controls and QMOFs, indicating systematic enrichment for higher multi-objective fitness.}
\label{fig:8}
\end{figure}

The control set consisted of nine experimentally established photocatalyst MOF, while the QMOF baseline comprised a random synthesized subset (n = 20).

Generated MOFs consistently outperformed curated photocatalyst controls and randomly sampled QMOF structures across the aggregated objective space. Dominance analysis showed complete separation between generated materials and controls, and near-complete separation between controls and QMOFs, indicating systematic enrichment for higher multi-objective fitness. Nonparametric statistical tests confirmed significant separation between all three tiers (Mann--Whitney U tests, p \textless{} 0.05 for all pairwise comparisons; Kruskal--Wallis test, p \textless{} 10⁻⁴), with large to very large effect sizes (Cohen's d = 1.79--2.16), demonstrating that observed differences are not only statistically significant but also practically meaningful. Full dominance fractions and statistical results are provided in the Supplementary Information.

\section{Discussion}

\subsection{Chemical Interpretation and Fitness Improvement}

As found in Fig.~\ref{fig:3}, post-hoc analysis of the top 1,000 candidates revealed that the N262 metal cluster (31.3\% prevalence) and BCG topology (34.0\%) emerged as recurring motifs, indicating that the model systematically identified chemically reasonable structural patterns. The Zn-based MOF\textquotesingle s N331 cluster (Zn center, N/C-rich framework) explains its strong sustainability and cost scores, while the Cr-based MOF\textquotesingle s N536 cluster (Cr center, sparse topology) balances high surface area with earth-abundant metal sourcing \cite{ref38}.

The model identifies candidates with an average 1.70±0.25-fold higher fitness score than existing materials, demonstrating its efficacy for multi-objective photocatalyst discovery. A Zn-based MOF with fitness score 1.20±0.05 fold higher than even the best material in the control set (PCN-224-Zr) was found. The workflow substantially reduces computational cost, from the introduction of ML utilized to explore the chemical space and to efficiently predict material properties to the funnel-based system improving computational efficiency by 4.13-fold (Supplemental Information).

\subsection{Multi-Objective Descriptor Trade-Off}

Our framework identified two Pareto-optimal MOF candidates representing distinct strategies for photocatalytic CO₂ reduction, with their normalized parameter scores detailed in Fig.~\ref{fig:5}a. The Zn-based MOF achieves maximum performance scores (catalytic: 1.00, adsorption: 1.00, stability: 1.00) at moderate economic cost (normalized cost: 0.44, sustainability: 0.13), positioning it as optimal for applications prioritizing conversion efficiency. Conversely, the Cr-based MOF sacrifices catalytic performance (0.52, falling 25\% below the PCN-224 benchmark) to achieve exceptional economic metrics (cost: 1.00, sustainability: 1.00), making it suitable for large-scale industrial deployment where lifecycle costs dominate. Critically, when economic factors were excluded ("No economic" function), Cr-MOF ranked below the PCN-224 benchmark (Z-score = -2.90, p = 0.004), demonstrating that its overall fitness superiority depends on cost and sustainability advantages rather than enhanced photocatalytic performance. Neither material simultaneously satisfies design criteria outlined in the introduction, confirming the fundamental performance-economics trade-off inherent to multi-objective catalyst optimization.

Parameter-level decomposition using Shapley value analysis (Fig.~\ref{fig:5}b) quantifies these trade-offs: Zn-MOF\textquotesingle s fitness improvement is driven by catalytic performance (30\% mean contribution across 13 fitness functions) and adsorption capacity (28\%), while Cr-MOF\textquotesingle s gain arises from economic advantages (sustainability: 67\%, cost: 36\%) that compensate for negative catalytic contribution (-2\%, indicating below-benchmark performance). This confirms that the two MOFs occupy fundamentally different regions of the multi-objective design space.

As shown in Fig.~\ref{fig:4}. The top candidates identified by the funnel satisfy multiple constraints simultaneously, indicating the workflow does not merely rank outliers but systematically uncovers structures that are well-balanced across mechanistic requirements. These predicted improvements align with established structure--property relationships in MOF photocatalysis, where chromium-centered frameworks exhibit enhanced charge separation and visible-light absorption \cite{ref56}, and Zn-based linkers optimize band-gap tuning and CO₂ activation \cite{ref57,ref58}.

\subsection{Uncertainty and Robustness Analysis}

Uncertainty propagation analysis (Fig.~\ref{fig:6}) validates these rankings under realistic prediction errors. At r = 0.11, representing typical CGCNN uncertainty \cite{ref17,ref31,ref48}, both candidates maintained statistical separation from PCN-224 across the majority of fitness functions: Zn-MOF satisfied the Z \textgreater{} 2 criterion in 11/13 functions (85\%), while Cr-MOF achieved 12/13 (92\%). Robustness persisted through r = 0.15 for both candidates (11/13 and 12/13 functions, respectively), demonstrating substantial safety margins beyond typical CGCNN prediction errors (5--12\%). At r = 0.20, representing severe uncertainty (±20\% relative error, approximately double typical prediction errors), the two MOFs exhibited divergent robustness profiles. While both maintained bounded improvements (lower bound \textgreater{} control upper bound) in most functions (Zn: 10/13, Cr: 12/13), the Z \textgreater{} 2 criterion revealed a critical distinction: Cr-MOF retained statistical significance in 10/13 functions (77\%), whereas Zn-MOF dropped to 0/13 (0\%).

Cr-MOF\textquotesingle s higher variability across functions (reflected in broader bootstrap confidence ranges: 1.31--1.88 vs. Zn\textquotesingle s 1.17--1.28) translates to larger propagated uncertainties, yet its extreme economic advantages (cost: 1.00, sustainability: 1.00) compared to the control create sufficiently large separation from the benchmark to maintain statistical significance even under severe perturbations \cite{ref43}. This divergence at r = 0.20 does not undermine confidence in either candidate, as both operate well within robust regimes under realistic uncertainty (r ≤ 0.15), and the r = 0.20 threshold primarily maps confidence limits rather than practical failure. At this extreme perturbation level, Cr-MOF retains Z \textgreater{} 2 in 77\% of functions, reflecting the robustness of its economic optimization profile, while Zn-MOF continues to show bounded improvements (10/13 functions) despite loss of formal statistical significance under unrealistically large error assumptions. For practical MOF screening applications operating at r ≈ 0.11, both candidates demonstrate equivalent and complete robustness.

\subsection{Photocatalytic Mechanistic Interpretation}
The predicted band-gap values for the top candidates, \textbf{1.907 eV for the Zn-based MOF} (cluster N331) and \textbf{1.735 eV for the Cr-based MOF} (cluster N536) fall within the visible-light regime commonly associated with experimentally active MOF photocatalysts \cite{ref59,ref60}, suggesting that their conduction and valence bands may reside within redox-accessible regions for CO₂ reduction and oxidative pathways \cite{ref61}. Although explicit DFT band-edge calculations were not performed, band gaps in the 1.8--2.0 eV range are frequently reported for Zn- and Cr-based frameworks capable of driving photoinduced redox processes, suggesting that the conduction- and valence-band positions of both structures may plausibly align with CO₂-reduction and oxidative pathways \cite{ref62,ref63,ref64}.

Although individual descriptors are computed through separate CGCNN models, the selected properties collectively capture mechanistic requirements for CO₂ photoreduction. The available surface area (ASA) governs the density of light-accessible adsorption sites and controls reactant diffusion to photoactive centers \cite{ref40,ref62}. The heat of adsorption influences stabilization of early CO₂RR intermediates such as *CO₂⁻ and *COOH \cite{ref65,ref66}, while CO₂/H₂O selectivity reduces competitive adsorption in humid flue-gas environments \cite{ref14,ref67}. Both top candidates surpass the control set in the adsorption composite metric, indicating a favorable balance between *CO₂ activation and product desorption. Water stability and bulk modulus are directly linked to hydrothermal durability and resistance to structural degradation during photocatalytic cycling \cite{ref14,ref33,ref34}. Cost and sustainability descriptors, though not mechanistic, determine scalability for industrial CO₂-remediation and oxidation-process reactors \cite{ref68,ref69}.

In addition, the enriched \textbf{bcg topology} and the recurring \textbf{N-containing coordination environment} in the Zn-based MOF (N331) are consistent with structural features that can support multidirectional charge migration in analogy to reported linker-mediated transport pathways in MOF photocatalysts \cite{ref70,ref71}. Independently of donor-atom identity, highly connected topologies such as bcg are widely associated with reduced electron--hole recombination due to increased transport dimensionality \cite{ref64,ref72,ref73}. These geometric and energetic considerations collectively support the photocatalytic plausibility of the MOFs prioritized by the funnel. Additional interpretation can be found in supplemental information S7.

From a mechanistic perspective, the selection of adsorption-related descriptors is motivated by the Sabatier principle, where optimal catalytic performance occurs at intermediate binding strengths of key intermediates \cite{ref74}. By explicitly optimizing adsorption energetics, CO₂/H₂O selectivity, and surface accessibility, the proposed ML framework identifies materials that reside in the favorable region of this Sabatier landscape, ensuring thermodynamic plausibility for photocatalytic CO₂ conversion prior to experimental validation.

\subsection{Lack of Explicit Photocatalytic Metrics}

While this work does not directly predict photocatalytic quantum yield (QY) or turnover frequency (TOF), the composite fitness functions used here integrate multiple mechanistic proxies known to correlate with photocatalytic performance. In practice, QY and TOF depend on complex, system-specific factors (e.g., photon flux, sacrificial agents, reaction pathways, charge carrier lifetimes) that are not directly accessible from structure-only databases such as QMOF. As a result, most large-scale computational screening studies rely on proxy descriptors and multi-objective fitness functions to prioritize candidates prior to detailed kinetic or experimental evaluation.

To assess whether the composite fitness functions capture meaningful photocatalytic structure--property relationships, we performed tiered validation against curated photocatalyst controls and randomly sampled QMOF structures (Fig.~\ref{fig:8}). Generated MOFs exhibited complete dominance over curated controls (dominance fraction = 1.00) and over random QMOFs (dominance fraction = 1.00), while curated controls also strongly outperformed QMOFs (dominance fraction = 0.967). Nonparametric tests confirmed statistically significant separation between all three tiers (Mann--Whitney U tests: p = 0.018, 4.1 × 10⁻⁵, and 0.0043 for Generated \textgreater{} Control, Control \textgreater{} QMOF, and Generated \textgreater{} QMOF, respectively; Kruskal--Wallis test: H = 19.19, p = 6.8 × 10⁻⁵), with large to very large effect sizes (Cohen's d = 1.79--2.16). This consistent hierarchical ordering (Generated \textgreater{} Control \textgreater{} QMOF) provides an internal construct-validity check, indicating that the fitness functions assign systematically higher scores to experimentally relevant photocatalysts than to random MOF structures and that the proposed descriptors and aggregation strategy are aligned with known photocatalytic trends.

\subsection{Proposed Preparation \& Feasibility of Identified MOFs}

Both top-ranked candidates exhibit close structural resemblance to experimentally reported MOFs in the QMOF database with their simulated XRD profiles (Fig.~\ref{fig:7}) align well with known crystallographic patterns \cite{ref17}. This agreement indicates physically reasonable atomistic connectivity and the absence of unphysical distortions \cite{ref75,ref76}. The metal-cluster motifs identified in these structures (N262, N331, and N536) correspond to coordination environments commonly found in well-established MOF families, including Zn--carboxylate frameworks \cite{ref77}, Cr-terephthalate systems such as MIL-101(Cr) \cite{ref78}, and N-rich linker architectures typical of porphyrin- and bipyridine-based MOFs \cite{ref79}. Because these families have reliable and widely demonstrated solvothermal synthesis routes, the structural motifs observed here fall firmly within chemically realistic and experimentally accessible coordination chemistry.

While no synthesis was performed in this study, literature precedent for Zn- and Cr-based frameworks suggests plausible solvothermal routes that are compatible with the coordination environments inferred from the identified motifs \cite{ref80}. Typical preparation pathways employ metal salts such as Zn(NO₃)₂·6H₂O, CrCl₃·6H₂O, or Cr(NO₃)₃·9H₂O in solvents including DMF, DEF, or ethanol--water mixtures \cite{ref78}. Reaction temperatures for related MOF families generally span 100--220 °C \cite{ref81}, with modulators such as benzoic acid or formic acid routinely used to control nucleation and crystal growth \cite{ref82}. These reactions are commonly conducted in Teflon-lined autoclaves at 20--50 mL laboratory scale \cite{ref77}. These conditions are not proposed as definitive synthesis routes for the predicted structures but illustrate that the linker environments and metal-cluster geometries identified by the model are consistent with established MOF chemistry. Consequently, the candidates generated lie within the domain of experimentally realizable frameworks rather than hypothetical constructs.

While Cr-based MOFs offer advantages in cost, abundance, and hydrothermal stability, the potential oxidation of Cr(III) to carcinogenic Cr(VI) under photo-oxidative conditions warrants careful consideration. Cr(VI) formation is primarily favored under strongly oxidative environments, high photon flux, and in the presence of reactive oxygen species, particularly in systems lacking strong ligand stabilization \cite{ref83}.

In MOF photocatalysts, chromium centers are typically stabilized within rigid octahedral Cr--O clusters, which kinetically hinder oxidation state changes \cite{ref84}. Moreover, many Cr-MOF photocatalysts operate under reductive CO₂ photoreduction conditions, which thermodynamically disfavors Cr(III) → Cr(VI) conversion \cite{ref85}. Strong metal--linker bonding and low defect densities have also been shown to suppress Cr leaching and oxidation \cite{ref86}.

If Cr-based MOFs are prioritized, toxicity risks can be further mitigated via surface passivation, redox-buffering strategies, and encapsulation, along with controlled reductive operation and routine oxidation-state monitoring (e.g., XPS, ICP-MS, or Cr(VI)-specific assays) \cite{ref87}. Thus, while chromium toxicity is a valid concern, it can be addressed through informed material design and operational control.

\subsection{Model Generalizability}

The proposed workflow generalizes naturally beyond CO₂ photoreduction to a wide range of advanced oxidation processes (AOPs), including photo-Fenton systems and related advanced oxidation processes that operate through visible-light--driven charge-transfer and radical-generation pathways \cite{ref88}. Many AOPs follow mechanistic steps analogous to CO₂RR: photoexcited electrons reduce dissolved oxygen to superoxide (O₂•⁻), while photogenerated holes oxidize water to generate hydroxyl radicals (•OH), leading to pollutant degradation \cite{ref89}. The descriptor set used in the funnel, adsorption metrics (ASA, HOA, CO₂/H₂O selectivity), water stability, bulk modulus, and band gap, maps directly onto these requirements \cite{ref64}. Adsorption descriptors are mechanism-agnostic and remain valid across CO₂RR and AOP environments, while water stability and mechanical robustness ensure structural integrity under hydrothermal and oxidative cycling \cite{ref16}. Band-gap predictions in the visible-light range support ROS formation, and high available surface area enhances interaction between pollutants or oxidants and photoreactive sites \cite{ref90}. Only the redox-oriented predictors (Faradaic efficiency, free energy, voltage potential) require domain-specific retraining for non-CO₂ pathways, which can be replaced with descriptors for ROS generation, band-edge alignment, or charge-carrier mobility \cite{ref41}. Because the funnel performs sequential multi-descriptor screening rather than optimizing a single catalytic metric, it captures trade-offs among adsorption, stability, energetics, and light absorption, allowing the same architecture to be repurposed efficiently for diverse photocatalytic material-discovery challenges.

\subsection{Research Limitations}

Multi-objective screening can be sensitive to fitness-function formulation. Improper weighting may bias selection toward overrepresented descriptors \cite{ref91,ref92}. To ensure robustness, three complementary validation strategies were applied:

(i) Pareto dominance analysis: The Zn-based MOF outperformed PCN-224(Zr) across all five normalized descriptors (Fig.~\ref{fig:5}a), confirming that its superiority holds under any monotonic aggregation function.

(ii) Multi-function ensemble: Rankings were evaluated across 13 alternative fitness formulations varying in mathematical form (e.g. additive, multiplicative, harmonic) and parameter emphasis (e.g. catalytic, economic, stability). Strong inter-function agreement (Kendall\textquotesingle s W = 0.79) confirms robustness to aggregation method choice (Fig.~\ref{fig:4}).

(iii) Uncertainty propagation: Fold improvements remained statistically significant across perturbation levels r = 0.05--0.20 (Fig.~\ref{fig:6}), demonstrating that ranking stability reflects genuine fitness advantages rather than CGCNN noise.

The Cr-based MOF, while not Pareto dominant, achieved top performance in 10 of 13 fitness formulations, reflecting the framework's ability to capture stability--economics trade-offs rather than overfitting to a single objective \cite{ref32,ref93}. The recurring N331 cluster and BCG topology across high-performing candidates further suggest chemically meaningful pattern recognition.

As with many computational screenings, experimental verification remains essential. The Zn- and Cr-based MOFs independently satisfy five orthogonal photocatalytic criteria: (i) synthesis feasibility (UFF convergence \cite{ref17}, synthesis accessibility score \textless{} 6, topological RMSD \textless{} 0.3 Å (Park et al., 2024), CGCNN synthesizability α \textgreater{} 0.5), (ii) predicted structural and aqueous stability, (iii) near optimal visible-light activity (band gap 1.73--2.0 eV) \cite{ref59,ref60}, (iv) balanced adsorption (ΔH\_ads = −20 to −60 kJ mol⁻¹; CO₂/H₂O selectivity) \cite{ref41,ref94}, and (v) experimentally consistent motifs (N262/N331 clusters, BCG topology); consistent XRD patterns, passed MOF-diff's matched connections criteria. The convergence of these independent constraints on the same two MOFs strengthens confidence in their predicted performance and indicates that high rankings are not artifacts of fitness function weighting. While these computational synthesizability indicators suggest high laboratory feasibility, predicted values remain unverified under photocatalytic conditions. Effects of photon flux, solvent, pH, and humidity could substantially influence true performance. Future work will focus on experimental synthesis and photocatalytic testing of the identified frameworks under visible-light irradiation to confirm CO₂ reduction activity and structural durability.

ML-guided MOF design frameworks do not explicitly consider vacancies or defect sites, which may overlook contributions from active defect centers \cite{ref95}. While vacancies can significantly influence photocatalytic activity, most ML-based approaches, including the present work, optimize properties such as adsorption, band gap, and stability without directly targeting defect engineering. Charge-carrier recombination diagnostics such as photoluminescence (PL) and electrochemical impedance spectroscopy (EIS) were not explicitly modeled here due to the absence of transferable structure-based predictors; their role and scope are discussed in the Supplementary Information.

\section{Conclusion}

Traditional MOF discovery for photocatalysis remains slow and largely focused on isolated material properties, with limited use of integrated machine-learning frameworks capable of simultaneously optimizing catalytic, adsorption, stability, and economic criteria across large chemical spaces. This study presents a machine-learning--guided framework for the rational discovery of photocatalytic metal--organic frameworks (MOFs) aimed at environmental remediation and CO₂ reduction. By combining reinforcement learning-based MOF generation with a multi-stage CGCNN prediction funnel, the workflow efficiently explored a broad structural space and identified materials exhibiting superior predicted photocatalytic performance under visible-light conditions. The top candidates, particularly Zn- and Cr-based MOFs, displayed enhanced adsorption characteristics, optimal band-gap ranges (1.8--2.0 eV), and high predicted stability in aqueous media, key requirements for advanced oxidation processes (AOPs) and pollutant degradation.

The proposed pipeline reduces computational cost by 4.13 fold while maintaining robustness across multiple performance descriptors, demonstrating that data-driven design can uncover MOFs with holistic improvements rather than isolated feature gains. Importantly, simulated X-ray diffraction confirmed strong structural similarity to known, synthesizable frameworks, supporting the practical feasibility of the generated materials.

Beyond CO₂ photoreduction, this modular methodology can be readily adapted to other AOP-relevant reactions such as the degradation of organic dyes, antibiotics, or volatile organics, by redefining the target descriptors in the predictive funnel. Future experimental validation will be crucial to verify photocatalytic activity, reaction kinetics, and long-term structural durability under real operating conditions. Overall, this work demonstrates that integrating machine learning with materials simulation provides a scalable and generalizable route to accelerate the discovery of efficient, stable, and sustainable MOF-based photocatalysts for environmental applications. The convergence of structural, electronic, adsorption, and stability descriptors indicates that the identified MOFs are not only theoretically promising but also practically viable starting points for synthesis and experimental evaluation.

\section*{Data Availability}

Training datasets used and trained CGCNN models can be found at our Zenodo: \href{https://doi.org/10.5281/zenodo.17581148}{10.5281/zenodo.17581148}. The base MOF-Optimized CGCNN architecture and supporting programs for workflow implementation can be found at our GitHub \url{https://github.com/SatyaK-0/MatCreatioNN}.

Data availability information can be found in Section S1 of this Supplementary Information.

\section*{Declaration of generative AI and AI-assisted technologies}

During the preparation of this work the author(s) used ChatGPT in order to improve language and check grammar. After using this tool/service, the author(s) reviewed and edited the content as needed and take(s) full responsibility for the content of the published article.

\clearpage
\section*{Supplementary Information}
\renewcommand{\thefigure}{S\arabic{figure}}
\setcounter{figure}{0}

\section*{S1 Code and Data Availability}

All code and datasets used in this study are publicly available at

Github: \url{https://github.com/SatyaK-0/MatCreatioNN/tree/main}

\begin{enumerate}
\def\labelenumi{\arabic{enumi}.}
\item
  MachineLearningPredictionModels\_PTHFiles/: The trained PyTorch Model
  File (.pth) of the 13 trained CGCNNs used in the funnel/filtration
  system.
\item
  LogicProof/: A short script demonstrating that the ordering of funnel
  stages does not affect final selection when only quantile-based
  filtering is used.
\item
  SupportingPythonPrograms\_Implementation\_DataProcessing/: --
  Supporting scripts for reinforcement-learning MOF generation,
  PyMatGen-based XRD simulation, molecular cost estimation, and all
  funnel-system post-processing routines.
\item
  cgcnn/: -- The modified and bug-fixed CGCNN implementation used in
  this work, updated for compatibility with current PyTorch and library
  versions.
\end{enumerate}

Zenodo: 10.5281/zenodo.17581148

\begin{enumerate}
\def\labelenumi{\arabic{enumi}.}
\item
  MOF\_MachineLearningDatasets/: The 8 MOF datasets used for machine
  learning/CGCNN training in the funnel system.
\item
  SupplementalDatasets/: The MOFSimplify Thermal and Water stability
  datasets, the MOF atomic-scale cost breakdown, the list of sustainable
  atoms as found in {[}1{]}. Additionally, a dataset of 381 MOFs with
  their respective experimental data was tabulated to assist future
  researchers in generalizability of this workflow to other AOP
  applications.
\end{enumerate}

\section*{S2 Reinforcement Learning}

The transformer-based reinforcement learning framework generates metal
clusters, organic linkers, and topologies to construct new MOFs. Each
generated MOF is tokenized into SMILES representations for its
structural components, and the decoder reconstructs the full framework.
The neural network predicts CO₂ heat of adsorption or CO₂/H₂O
selectivity, and its output serves as the reward signal guiding the
reinforcement learning process toward higher-performing MOFs {[}2{]}.

\section*{S3 Crystal Graph Neural Network Funnel System}

\subsection*{S3.1 Funnel Sensitivity}

The funnel applies sequential 5\% bottom-quantile cutoffs to reduce
low-performing candidates. Although alternative thresholds were not
exhaustively evaluated, the strong recurrence of bcg topology and
N262/N331/N536 clusters suggests that enrichment patterns are unlikely
to be sensitive to moderate threshold variation.

Topological motifs enriched by generative models tend to remain robust
under threshold variation because they reflect consistent alignment
between structural families and descriptor targets. The large initial
dataset (120,000 MOFs) further reduces sensitivity to small changes in
filtering parameters, since multiple independent candidates converge
onto similar motif families before filtration.

Thus, while only a 5\% cutoff is used here, the interpretability of
post-hoc motif enrichment provides qualitative evidence that the
pipeline does not depend on a single arbitrary threshold.

\subsection*{S3.2 Pareto-Front-Based Hyperparameter Selection}

\textbf{Rationale}

Model performance in catalyst prediction involves two competing
objectives: \textbf{predictive accuracy} (Test MAE) and
\textbf{computational efficiency} (Test Time).\\
A single scalar metric (e.g., MAE alone) cannot capture this trade-off.
Therefore, a \textbf{Pareto front} was constructed to identify
non-dominated models---those not outperformed simultaneously in both
accuracy and speed.

To ensure comparability, both objectives were normalized between 0 and 1
using min--max scaling:

\[x^{'} = \frac{x - \min(x)}{\max(x) - \min(x)}
\]

The Pareto front was obtained by sorting models by increasing Test MAE
and retaining those with monotonically decreasing Test Time.

Subsequently, a \textbf{weighted Euclidean distance} from the ideal
point (0, 0) was used to select the most balanced configuration:

\[D = \sqrt{0.76(\text{MAE}^{'})^{2} + 0.24(\text{Time}^{'})^{2}}
\]

Here, MAE was given 76\% importance, reflecting the higher scientific
value of predictive fidelity for photocatalyst screening relative to
small differences in computational time.

\begin{figure}[htbp]
\centering
\includegraphics[width=0.9\linewidth]{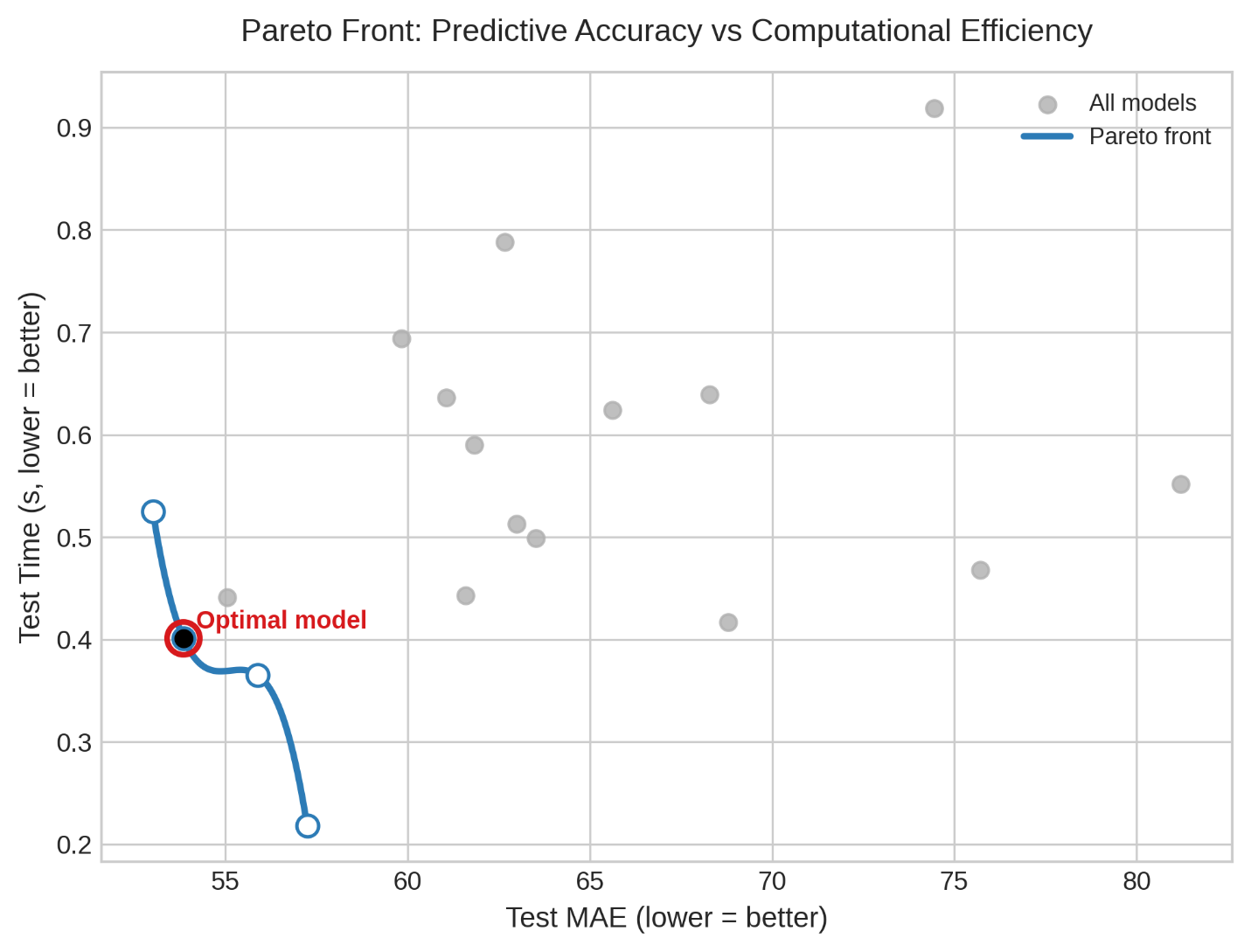}
\caption{Pareto front illustrates the trade-off between
predictive accuracy (Test MAE) and computational efficiency (Test Time)
across all tested configurations of the MOF-photocatalyst prediction
model. Gray circles denote all evaluated models, blue open circles mark
Pareto-optimal points, and the blue dashed line traces the continuous
Pareto frontier. The selected optimal model (red-outlined circle, black
core) minimizes a weighted Euclidean distance D above.}
\label{fig:s1}
\end{figure}

The resulting Pareto front reveals a monotonic trade-off between
\textbf{model accuracy} and \textbf{execution time}.\\
The lower-left boundary represents configurations offering the best
achievable accuracy for a given computational cost.\\
Applying a 0.76:0.24 accuracy--time weighting identified the
configuration (Test MAE = 53.852, Test Time = 0.401 s) as the global
optimum, highlighted in Fig.~\ref{fig:s1}.\\
This balance prioritizes reliable prediction of photocatalytic activity
while retaining computational tractability, aligning with the
framework's goal of scalable materials screening.

\begin{longtable}[]{@{}
  >{\raggedright\arraybackslash}p{(\columnwidth - 8\tabcolsep) * \real{0.3184}}
  >{\raggedright\arraybackslash}p{(\columnwidth - 8\tabcolsep) * \real{0.1762}}
  >{\raggedright\arraybackslash}p{(\columnwidth - 8\tabcolsep) * \real{0.2424}}
  >{\raggedright\arraybackslash}p{(\columnwidth - 8\tabcolsep) * \real{0.1057}}
  >{\raggedright\arraybackslash}p{(\columnwidth - 8\tabcolsep) * \real{0.1573}}@{}}
\toprule\noalign{}
\begin{minipage}[b]{\linewidth}\raggedright
Learning Rate
\end{minipage} & \begin{minipage}[b]{\linewidth}\raggedright
n-conv
\end{minipage} & \begin{minipage}[b]{\linewidth}\raggedright
Batch size
\end{minipage} & \begin{minipage}[b]{\linewidth}\raggedright
n-h
\end{minipage} & \begin{minipage}[b]{\linewidth}\raggedright
epoch
\end{minipage} \\
\midrule\noalign{}
\endhead
\bottomrule\noalign{}
\endlastfoot
0.01 & 4 & 32 & 3 & 30 \\
\end{longtable}

\par\noindent\textbf{Table S1:} Finalized CGCNN hyperparameters selected from the
Pareto-optimized grid search, showing the learning rate, number of
convolutional layers (n-conv), batch size, number of hidden layers
(n-h), and total epochs used for training

\subsection*{S3.2 Computational Modelling Proxy.}

To ensure clarity and reproducibility, we provide the exact mathematical
forms of all proxy functions used to approximate catalytic behavior
within the GAPF framework.

The Gaussian weighting function applied to predicted band-gap values,
\textbf{B}, is:

\[f(B) = {4.536*e}^{(\frac{- (B - 1.9)^{2}}{2*75}}\]

For the adsorption descriptor \textbf{A}, a truncated Gaussian weighting
function is defined as:

\[g(A) = e^{\frac{- (A + 30)^{2}}{2*75}}\]

and its truncated (bounded) form is:

\[h(A) = \frac{1}{g( - 40)}*min(\ g(x),\ g( - 40))\]

\subsection*{S3.3 Cost and Sustainability Dataset Creation}

To incorporate economic and environmental constraints into the funnel,
two domain-specific CGCNN predictors were constructed: one for material
cost and one for sustainability. The cost dataset was generated by
calculating the elemental cost of each MOF using a curated table of
element prices in their common industrial forms (MOFCreatioNN GitHub;
SatyaK-0, 2024). Each MOF's total cost was computed as the sum of its
constituent atomic costs.

C\_MOF = Σ(n\_i × p\_i)

where n\_i is the count of element i in the MOF formula unit and p\_i is
the price per mole of element. Complete price tables and calculation
scripts are available at github (S1)

Sustainability scores were defined at the atomic level based on whether
each element could be sourced from industrial waste or post-consumer
waste streams, following the classification scheme of {[}1{]}. Each atom
contributing ``1'' to the sustainability count was normalized by the
total atom count of the MOF. Because raw scores were extremely small
(mean ≈ 0.0021 across QMOF), all sustainability values were multiplied
by 1000 to maintain numerical stability during CGCNN training and to
avoid vanishing gradients. The sustainably scored atoms are available at
the GitHub (S1).

The QMOF dataset is highly imbalanced with respect to sustainability,
22,609 of 23,750 structures contained no sustainable atoms. To counter
this imbalance, the training set was constructed using random
undersampling (U\_Random), selecting all 1,141 MOFs with nonzero
sustainability and pairing them with 1,141 randomly selected
zero-sustainability MOFs, yielding a balanced training set of 2,282
structures. U\_random was shown to be the best undersampling technique
for unbalanced datasets {[}3{]}. This procedure follows recommended
practice for unbalanced materials datasets and ensures that the
sustainability predictor is sensitive to the presence of waste-derived
metals.

\subsection*{S3.4 CGCNN Training Results}

The crystal graph convolutional neural network (CGCNN) models used
within the GAPF framework were trained following standard hyperparameter
as mentioned in section S1 and normalization protocols established in
prior literature. To ensure transparency and reproducibility, we report
the dataset origin, train/validation/test partitioning, and predictive
accuracy for each property-specific model used in the funnel.

\begin{longtable}[]{@{}
  >{\raggedright\arraybackslash}p{(\columnwidth - 10\tabcolsep) * \real{0.2056}}
  >{\raggedright\arraybackslash}p{(\columnwidth - 10\tabcolsep) * \real{0.2253}}
  >{\raggedright\arraybackslash}p{(\columnwidth - 10\tabcolsep) * \real{0.1396}}
  >{\raggedright\arraybackslash}p{(\columnwidth - 10\tabcolsep) * \real{0.1396}}
  >{\raggedright\arraybackslash}p{(\columnwidth - 10\tabcolsep) * \real{0.1646}}
  >{\raggedright\arraybackslash}p{(\columnwidth - 10\tabcolsep) * \real{0.1253}}@{}}
\toprule\noalign{}
\begin{minipage}[b]{\linewidth}\raggedright
Predictor
\end{minipage} & \begin{minipage}[b]{\linewidth}\raggedright
Dataset origin
\end{minipage} & \begin{minipage}[b]{\linewidth}\raggedright
Total Dataset size
\end{minipage} & \begin{minipage}[b]{\linewidth}\raggedright
Training Set Size
\end{minipage} & \begin{minipage}[b]{\linewidth}\raggedright
Validation Set Size
\end{minipage} & \begin{minipage}[b]{\linewidth}\raggedright
Testing Set Size
\end{minipage} \\
\midrule\noalign{}
\endhead
\bottomrule\noalign{}
\endlastfoot
Water Stability¹ & MOFSimplify {[}4{]} & 36 x 10 & \textasciitilde28 &
\textasciitilde4 & \textasciitilde4 \\
Cost² & QMOF {[}5{]} & 20374 & 16300 & 2037 & 2037 \\
Sustainability & QMOF {[}5{]} & 2282 & 1718 & 282 & 282 \\
Temperature & MOFSimplify {[}4{]} & 3168 & 2534 & 317 & 317 \\
Selectivity & MOFReinforce {[}2{]} & 3980 & 3184 & 398 & 398 \\
Heat of Adsorption & MOFReinforce {[}2{]} & 5188 & 4150 & 519 & 519 \\
Available Surface Area & MOSAEC {[}6{]} & 19962 & 15970 & 1996 & 1996 \\
Synthesis Predictor & QMOF {[}5{]} & 20375 & 16299 & 2038 & 2038 \\
\end{longtable}

\par\noindent\textbf{Table S2:} Dataset sources and data partitioning for all CGCNN property
predictors. Each predictor was trained using the dataset listed, with
80/10/10 splits applied unless noted otherwise. Dataset origins include
MOFSimplify, QMOF, MOFReinforce, and MOSAEC.

¹Given small dataset size, 10-fold cross validation was conducted. As
such, some folds used slightly varied Training/Validation/Testing Size.

²A MOF was invalid as input to our cost preprocessor and so the
datapoint was removed.

\begin{longtable}[]{@{}
  >{\raggedright\arraybackslash}p{(\columnwidth - 8\tabcolsep) * \real{0.1658}}
  >{\raggedright\arraybackslash}p{(\columnwidth - 8\tabcolsep) * \real{0.2222}}
  >{\raggedright\arraybackslash}p{(\columnwidth - 8\tabcolsep) * \real{0.2137}}
  >{\raggedright\arraybackslash}p{(\columnwidth - 8\tabcolsep) * \real{0.2040}}
  >{\raggedright\arraybackslash}p{(\columnwidth - 8\tabcolsep) * \real{0.1943}}@{}}
\toprule\noalign{}
\begin{minipage}[b]{\linewidth}\raggedright
Predictor
\end{minipage} & \begin{minipage}[b]{\linewidth}\raggedright
Train MAE
\end{minipage} & \begin{minipage}[b]{\linewidth}\raggedright
Val MAE
\end{minipage} & \begin{minipage}[b]{\linewidth}\raggedright
Test MAE
\end{minipage} & \begin{minipage}[b]{\linewidth}\raggedright
Test RMSE
\end{minipage} \\
\midrule\noalign{}
\endhead
\bottomrule\noalign{}
\endlastfoot
Bulk Moduli³ & Not Reported & Not Reported & 0.054 log(GPa) & Not
Reported \\
Water Stability & 0.027 & 0.064 & 0.084 & 0.112 \\
Band Gap³ & Not Reported & Not Reported & 0.388 eV & Not Reported \\
Cost & 0.005 & 0.035 & 0.03 & 0.08 \\
Sustainability & 0.007 & 0.008 & 0.008 & 0.026 \\
Temperature & 0.002 & 0.08 & 0.08 & 0.11 \\
Selectivity & 0.029 & 0.086 & 0.088 & 0.13 \\
Heat of Adsorption & 0.004 & 0.056 & 0.055 & 0.084 \\
Available Surface Area & 0.070 & 0.076 & 0.076 & 0.101 \\
\end{longtable}

\par\noindent\textbf{Table S3:} Summary of predictive accuracy for all CGCNN models used in
the funnel, reporting training, validation, and test MAE values, along
with test RMSE and R² where available. Predictors sourced from external
work (bulk modulus and band gap) are reported using their published
metrics. All properties are normalized between {[}0-1{]} for scaling.

³Full regression training, predicted vs. actual scatter plot and other
training/validation data can be found at {[}7{]}. They do not publicly
report on the other training, validation or testing metrics.

\begin{longtable}[]{@{}
  >{\raggedright\arraybackslash}p{(\columnwidth - 14\tabcolsep) * \real{0.1296}}
  >{\raggedright\arraybackslash}p{(\columnwidth - 14\tabcolsep) * \real{0.1682}}
  >{\raggedright\arraybackslash}p{(\columnwidth - 14\tabcolsep) * \real{0.1733}}
  >{\raggedright\arraybackslash}p{(\columnwidth - 14\tabcolsep) * \real{0.1059}}
  >{\raggedright\arraybackslash}p{(\columnwidth - 14\tabcolsep) * \real{0.1398}}
  >{\raggedright\arraybackslash}p{(\columnwidth - 14\tabcolsep) * \real{0.1215}}
  >{\raggedright\arraybackslash}p{(\columnwidth - 14\tabcolsep) * \real{0.0809}}
  >{\raggedright\arraybackslash}p{(\columnwidth - 14\tabcolsep) * \real{0.0809}}@{}}
\toprule\noalign{}
\begin{minipage}[b]{\linewidth}\raggedright
\end{minipage} & \begin{minipage}[b]{\linewidth}\raggedright
Train F1
\end{minipage} & \begin{minipage}[b]{\linewidth}\raggedright
Train Cross-Entropy Loss
\end{minipage} & \begin{minipage}[b]{\linewidth}\raggedright
Val F1
\end{minipage} & \begin{minipage}[b]{\linewidth}\raggedright
Val Cross-Entropy Loss
\end{minipage} & \begin{minipage}[b]{\linewidth}\raggedright
Test Accuracy
\end{minipage} & \begin{minipage}[b]{\linewidth}\raggedright
Test F1
\end{minipage} & \begin{minipage}[b]{\linewidth}\raggedright
Test AUC
\end{minipage} \\
\midrule\noalign{}
\endhead
\bottomrule\noalign{}
\endlastfoot
Synthesis Predictor & 0.99 & 0.007 & 0.99 & 0.99 & 0.99 & 0.99 & 0.99 \\
\end{longtable}

\par\noindent\textbf{Table S4:} Classification performance of the CGCNN classification
predictor. Reported metrics include training and validation F1 scores
and cross-entropy losses, along with test-set accuracy, F1 score, and
area under the ROC curve (AUC)

For:

\begin{enumerate}
\def\labelenumi{\arabic{enumi}.}
\item
  Faradaic Efficiency, Free Energy, Voltage Potential
\end{enumerate}

And their regression training, predicted vs. actual scatter plot and
other training/validation data can be found at {[}8{]}.

\subsection*{S3.5 Computational Cost Comparison}

In reporting the computational cost of the screening workflow, we
quantify performance in terms of the number of model inferences required
rather than wall-clock runtimes. Absolute runtimes depend heavily on
implementation details, hardware variability, parallelization settings,
GPU/CPU availability, and background load on shared compute clusters,
making them unreliable or non-reproducible across different
environments. In contrast, the number of forward passes each model must
perform is an architecture-dependent quantity that remains stable
regardless of hardware. Reporting computational cost in units of
inference counts therefore provides a hardware-agnostic, reproducible
measure of algorithmic complexity and allows other researchers to
estimate their own expected runtimes based on local compute resources.
This approach aligns with best practices in machine-learning
benchmarking, where inference operations, not raw wall-clock times, are
treated as the primary measure of computational demand {[}9{]}.

Standard High-Throughput Search Methodology

120,000 MOFs * 13 property predictions = 1,560,000 inferences.

Proposed funnel system:

Post preliminary filter

\begin{longtable}[]{@{}
  >{\raggedright\arraybackslash}p{(\columnwidth - 6\tabcolsep) * \real{0.2342}}
  >{\raggedright\arraybackslash}p{(\columnwidth - 6\tabcolsep) * \real{0.2769}}
  >{\raggedright\arraybackslash}p{(\columnwidth - 6\tabcolsep) * \real{0.2547}}
  >{\raggedright\arraybackslash}p{(\columnwidth - 6\tabcolsep) * \real{0.2342}}@{}}
\toprule\noalign{}
\begin{minipage}[b]{\linewidth}\raggedright
Stage
\end{minipage} & \begin{minipage}[b]{\linewidth}\raggedright
Computation
\end{minipage} & \begin{minipage}[b]{\linewidth}\raggedright
Inferences at Stage
\end{minipage} & \begin{minipage}[b]{\linewidth}\raggedright
Total Inferences
\end{minipage} \\
\midrule\noalign{}
\endhead
\bottomrule\noalign{}
\endlastfoot
Stability & 35352 MOFs * 2 predictors & 70704 & 70704 \\
Catalytic Ability & 33582 MOFs * 4 predictors & 134328 & 205032 \\
Cost & 31902 MOFs * 1 predictor & 31902 & 236934 \\
Sustainability & 30307 MOFs * 1 predictor & 30307 & 267241 \\
Temperature Stability & 27391 MOFs * 1 predictor & 27391 & 294632 \\
Adsorption & 20943 MOFs * 3 predictors & 62829 & 357461 \\
Synthesis Predictor & 19876 MOFs * 1 predictor & 19876 & 377337 \\
\end{longtable}

\par\noindent\textbf{Table S5:} Computational cost breakdown of the CGCNN funnel system. For
each stage, the number of MOFs evaluated, and the corresponding number
of model inferences are listed, along with cumulative totals. This
quantifies the efficiency gained by staged filtering.

This proposed funnel had used 377,337 inferences as compared to the
high-throughput search methodology of 1,560,000. A 4.13-fold
improvement.

\section*{S4 Material Benchmark}

To benchmark the results and materials identified, a control set of the
most promising photocatalysts from various review articles were
identified. The generated MOFs were compared against existing materials
by running all the control MOF set through the same CGCNN module-
utilizing the same post-processing and fitness function to determine
their strength.

Machine learning models have optimal performance with minimal
extrapolation from the training set, as to maintain accurate prediction,
we kept the datasets used for CGCNN training for a MOF-focus. As such,
to maintain consistency when benchmarking, the benchmarks must also be
solely MOF photocatalysts as well.

\begin{longtable}[]{@{}
  >{\raggedright\arraybackslash}p{(\columnwidth - 2\tabcolsep) * \real{0.5012}}
  >{\raggedright\arraybackslash}p{(\columnwidth - 2\tabcolsep) * \real{0.4988}}@{}}
\toprule\noalign{}
\endhead
\bottomrule\noalign{}
\endlastfoot
MOF Name & Source detailing MOF photocatalytic ability \\
MOF-74 (Co) & {[}10{]} \\
UiO-66(Zr) & {[}11{]} \\
PCN-222 & {[}12{]} \\
PCN-224 & {[}13{]} \\
MIL-101 (Cr) & {[}14{]} \\
MIL-53-NH2 (Fe)\textsuperscript{a} & {[}15{]} \\
MOF-808 & {[}16{]} \\
\end{longtable}

\par\noindent\textbf{Table S6:} Literature sources documenting photocatalytic activity for the
benchmark MOFs evaluated in this study. These references confirm that
each control structure has demonstrated experimental photocatalytic or
CO₂-reduction performance under visible-light or related photochemical
conditions.

\section*{S5 Fitness Function Proxy}

\textbf{S5.1 Ensemble Fitness Function}

To evaluate robustness of the scoring framework, we compared a series of
alternative fitness formulations that combine normalized material
descriptors, stability (\(S_{\text{stab}}\)), catalytic activity
(\(S_{\text{cat}}\)), cost (\(S_{\text{cost}}\)), sustainability
(\(S_{\text{sust}}\)), and adsorption efficiency (\(S_{\text{ads}}\)).\\
All parameters were scaled to the range {[}0, 1{]} using min--max
normalization; the cost descriptor was inverted such that lower raw cost
corresponds to a higher normalized score.

Each input
\(x_{i} \in \{ S,C_{\text{cat}},C_{\text{cost}},C_{\text{sust}},A\}\)is
assigned a \textbf{conservative uniform relative uncertainty} of
\(r = 11\%\). Uncertainty in each composite fitness score arises from
the propagation of relative uncertainties in its five normalized input
descriptors:

\[\mathbf{x} = (S,C_{\text{cat}},C_{\text{cost}},C_{\text{sust}},A)
\]

where each represents a normalized property (stability, catalytic
activity, economic cost, sustainability, adsorption capacity).

Thus, the absolute standard deviation is
\(\sigma_{x_{i}} = r\text{ }x_{i}\). Inputs are treated as independent
for first order (Gaussian) error propagation:

\[\sigma_{f}^{2} \approx \sum_{i}^{}{}{(\frac{\partial f}{\partial x_{i}})}^{2}\sigma_{x_{i}}^{2}.
\]

Unless noted, all logarithms are natural logs, and all results are
reported as \textbf{mean} \(\pm\)\textbf{1-σ}.

Let
\(\mathbf{x} = (S,C_{\text{cat}},C_{\text{cost}},C_{\text{sust}},A)\).

\textbf{Eq. S1 (Additive, equal weights).}

\[f_{1} = S + C_{\text{cat}} + C_{\text{cost}} + C_{\text{sust}} + A
\]

Propagation:

\[\sigma_{f_{1}} = \sqrt{\sum_{i}^{}(rx_{i})^{2}}.
\]

\textbf{Eq. S2 (Multiplicative, equal---geometric mean).}

\[f_{2} = (S\text{ }C_{\text{cat}}\text{ }C_{\text{cost}}\text{ }C_{\text{sust}}\text{ }A)^{1/5}
\]

Using \(\ln f_{2} = \frac{1}{5}\sum_{i}^{}{}\ln x_{i}\), the relative
variance is

\[{(\frac{\sigma_{f_{2}}}{f_{2}})}^{2} = \frac{1}{25}\sum_{i}^{}{}{(\frac{\sigma_{x_{i}}}{x_{i}})}^{2} = \frac{5\text{ }r^{2}}{25} = \frac{r^{2}}{5}.
\]

Hence \(\sigma_{f_{2}} = f_{2}\text{ }r/\sqrt{5}\)(value-independent
factor for equal \(r\)).

\textbf{Eq. S3 (Catalytic emphasis, arithmetic).}

\[f_{3} = \frac{S + 2C_{\text{cat}} + C_{\text{cost}} + C_{\text{sust}} + A}{6},\sigma_{f_{3}} = \sqrt{\sum_{i}^{}{(\frac{w_{i}}{6}\text{ }r\text{ }x_{i})}^{2}},\text{ }w = (1,2,1,1,1).
\]

\textbf{Eq. S4 (Economic emphasis, arithmetic).}

\[f_{4} = \frac{S + C_{\text{cat}} + 2C_{\text{cost}} + 2C_{\text{sust}} + A}{7},\sigma_{f_{4}} = \sqrt{\sum_{i}^{}{(\frac{w_{i}}{7}\text{ }r\text{ }x_{i})}^{2}},\text{ }w = (1,1,2,2,1).
\]

\textbf{Eq. S5 (Stability \& adsorption emphasis, arithmetic).}

\[f_{5} = \frac{2S + C_{\text{cat}} + C_{\text{cost}} + C_{\text{sust}} + 2A}{7},\sigma_{f_{5}} = \sqrt{\sum_{i}^{}{(\frac{w_{i}}{7}\text{ }r\text{ }x_{i})}^{2}},\text{ }w = (2,1,1,1,2).
\]

\textbf{Eq. S6 (Harmonic mean of five).}

\[f_{6} = \frac{5}{\sum_{i}^{}1/x_{i}},\frac{\partial f_{6}}{\partial x_{i}} = \frac{5}{{(\sum_{j}^{}1/x_{j})}^{2}} \cdot \frac{1}{x_{i}^{2}}.
\]

Thus

\[\sigma_{f_{6}} = \sqrt{\sum_{i}^{}{\lbrack\frac{5}{{(\sum_{j}^{}1/x_{j})}^{2}} \cdot \frac{1}{x_{i}^{2}} \cdot rx_{i}\rbrack}^{2}}.
\]

\textbf{Eq. S7 (Geometric, catalytic squared; 6th-root).}

\[f_{7} = (S\text{ }C_{\text{cat}}^{2}\text{ }C_{\text{cost}}\text{ }C_{\text{sust}}\text{ }A)^{1/6},\text{ }{(\frac{\sigma_{f_{7}}}{f_{7}})}^{2} = \frac{1}{36}(1^{2} + 2^{2} + 1^{2} + 1^{2} + 1^{2})r^{2} = \frac{8}{36}r^{2}.
\]

Hence \(\sigma_{f_{7}} = f_{7}\text{ }r\text{ }\sqrt{8}/6\).

\textbf{Eq. S8 (Min--max average).}

\[f_{8} = \frac{\min(\mathbf{x}) + \max(\mathbf{x})}{2}.
\]

Non-differentiable at order switches; we use a conservative bound where
only the active \(\min\)and \(\max\)contribute:

\[\sigma_{f_{8}} = \frac{1}{2}\sqrt{(r\text{ }\min)^{2} + (r\text{ }\max)^{2}}.
\]

\textbf{Eq. S9 (Multiplicative without catalytic; 4th-root).}

\[f_{9} = (S\text{ }C_{\text{cost}}\text{ }C_{\text{sust}}\text{ }A)^{1/4},{(\frac{\sigma_{f_{9}}}{f_{9}})}^{2} = \frac{1}{16} \cdot 4\text{ }r^{2} \Rightarrow \sigma_{f_{9}} = f_{9}\text{ }r/2.
\]

\textbf{Eq. S10 (Additive without economics; arithmetic mean of 3).}

\[f_{10} = \frac{S + C_{\text{cat}} + A}{3},\sigma_{f_{10}} = \sqrt{\sum_{x \in \{ S,C_{\text{cat}},A\}}^{}{(\frac{r\text{ }x}{3})}^{2}}.
\]

\textbf{Eq. S11 (Square-root of sum).}

\[{f_{11} = \sqrt{S + C_{\text{cat}} + C_{\text{cost}} + C_{\text{sust}} + A},\frac{\partial f_{11}}{\partial x_{i}} = \frac{1}{2\sqrt{\sum_{j}^{}x_{j}}},
}{\sigma_{f_{11}} = \sqrt{\sum_{i}^{}{(\frac{r\text{ }x_{i}}{2\sqrt{\sum_{j}^{}x_{j}}})}^{2}}.
}\]

\textbf{Eq. S12 (Log-sum ``product'' proxy).}

\[{f_{12} = \sum_{i}^{}\ln(x_{i} + 0.1),\frac{\partial f_{12}}{\partial x_{i}} = \frac{1}{x_{i} + 0.1},
}{\sigma_{f_{12}} = \sqrt{\sum_{i}^{}{(\frac{r\text{ }x_{i}}{x_{i} + 0.1})}^{2}}.
}\]

\textbf{Eq. S13 (Extreme economic average).}

\[f_{13} = \frac{C_{\text{cost}} + C_{\text{sust}}}{2},\sigma_{f_{13}} = \sqrt{{(\frac{r\text{ }C_{\text{cost}}}{2})}^{2} + {(\frac{r\text{ }C_{\text{sust}}}{2})}^{2}}.
\]

\textbf{S5.2. Expert Weighted Fitness Function}

To complement the ensemble of thirteen fitness formulations, we
introduce a domain-informed composite metric that reflects mechanistic
priorities specific to photocatalytic CO₂ reduction. While the ensemble
analysis evaluates robustness across mathematically diverse
aggregations, practical catalyst selection often benefits from a single
expert-weighted function incorporating established structure--property
relationships. The expert fitness function below was therefore
constructed to prioritize (i) structural and hydrothermal stability,
(ii) catalytic activity relevant to CO₂ activation, and (iii)
adsorption-driven accessibility of photoreactive sites, while
incorporating sustainability and cost as secondary, but practically
important, constraints. This function is used exclusively for post hoc
interpretation of the candidate design space and material motifs rather
than for filtering within the funnel.

\[\frac{E_{\text{stab}}*E_{\text{cat}}*E_{\text{ads}}*\sqrt{E_{\text{sust}} + 1}}{\sqrt{E_{\text{cost}} + 1}}\]

\section*{S6 Promising Material Identification and Parameter}

\textbf{S6.1 Raw and Normalized Parameter Values}

\begin{longtable}[]{@{}
  >{\raggedright\arraybackslash}p{(\columnwidth - 8\tabcolsep) * \real{0.2329}}
  >{\raggedright\arraybackslash}p{(\columnwidth - 8\tabcolsep) * \real{0.1941}}
  >{\raggedright\arraybackslash}p{(\columnwidth - 8\tabcolsep) * \real{0.1669}}
  >{\raggedright\arraybackslash}p{(\columnwidth - 8\tabcolsep) * \real{0.2700}}
  >{\raggedright\arraybackslash}p{(\columnwidth - 8\tabcolsep) * \real{0.1360}}@{}}
\toprule\noalign{}
\begin{minipage}[b]{\linewidth}\raggedright
\textbf{MOF}
\end{minipage} & \begin{minipage}[b]{\linewidth}\raggedright
\textbf{Parameter}
\end{minipage} & \begin{minipage}[b]{\linewidth}\raggedright
\textbf{Raw Value}
\end{minipage} & \begin{minipage}[b]{\linewidth}\raggedright
\textbf{Normalized Value}
\end{minipage} & \begin{minipage}[b]{\linewidth}\raggedright
\textbf{Units}
\end{minipage} \\
\midrule\noalign{}
\endhead
\bottomrule\noalign{}
\endlastfoot
\textbf{PCN-224(Zr)} & Stability & 2.054 & 0.885 & score \\
& Catalytic & 17.303 & 0.689 & score \\
& Cost & 1498.32 & 0.381 & \$/kg \\
& Sustainability & 3.075 & 0.103 & score \\
& Adsorption & 1.630 & 0.727 & score \\
\textbf{Zn-based MOF} & Stability & 2.322 & \textbf{1.000} & score \\
& Catalytic & 25.098 & \textbf{1.000} & score \\
& Cost & 1313.60 & 0.435 & \$/kg \\
& Sustainability & 3.916 & 0.132 & score \\
& Adsorption & 2.242 & \textbf{1.000} & score \\
\textbf{Cr-based MOF} & Stability & 2.005 & 0.863 & score \\
& Catalytic & 12.919 & 0.515 & score \\
& Cost & 570.83 & \textbf{1.000} & \$/kg \\
& Sustainability & 29.760 & \textbf{1.000} & score \\
& Adsorption & 1.704 & 0.760 & score \\
\end{longtable}

\par\noindent\textbf{Table S7:} Full parameter dataset for PCN-224(Zr), the Zn-based MOF, and
the Cr-based MOF. Raw descriptor values and their normalized
counterparts are shown for all five composite parameters (stability,
catalytic ability, cost, sustainability, and adsorption).

\textbf{Critical observation}: Neither material achieves Pareto-optimal
status.

\textbf{Zn-based MOF profile:}

\begin{itemize}
\item
  \textbf{Strengths}: Stability (1.00), Catalytic (1.00), Adsorption
  (1.00)
\item
  \textbf{Weaknesses}: Cost (0.435), Sustainability (0.132)
\item
  \textbf{Character}: Catalytic performance leader with moderate
  economic profile
\end{itemize}

\textbf{Cr-based MOF profile:}

\begin{itemize}
\item
  \textbf{Strengths}: Cost (1.00), Sustainability (1.00)
\item
  \textbf{Weaknesses}: Catalytic (0.515), Stability (0.863), Adsorption
  (0.760)
\item
  \textbf{Character}: Economic/sustainability leader with compromised
  catalytic performance
\end{itemize}

\textbf{Pareto dominance analysis:}

\begin{itemize}
\item
  Zn-MOF dominates PCN-224 in: 4/5 parameters (all except
  sustainability)
\item
  Cr-MOF dominates PCN-224 in: 3/5 parameters (cost, sustainability,
  adsorption)
\item
  \textbf{Zn vs. Cr:} Neither dominates - each wins 2-3 parameters
\end{itemize}

This establishes a \textbf{genuine multi-objective trade-off} rather
than one-sided superiority.

\textbf{S6.2 Parameter Range Analysis}

\begin{longtable}[]{@{}
  >{\raggedright\arraybackslash}p{(\columnwidth - 8\tabcolsep) * \real{0.2030}}
  >{\raggedright\arraybackslash}p{(\columnwidth - 8\tabcolsep) * \real{0.1611}}
  >{\raggedright\arraybackslash}p{(\columnwidth - 8\tabcolsep) * \real{0.1669}}
  >{\raggedright\arraybackslash}p{(\columnwidth - 8\tabcolsep) * \real{0.1048}}
  >{\raggedright\arraybackslash}p{(\columnwidth - 8\tabcolsep) * \real{0.3641}}@{}}
\toprule\noalign{}
\begin{minipage}[b]{\linewidth}\raggedright
\textbf{Parameter}
\end{minipage} & \begin{minipage}[b]{\linewidth}\raggedright
\textbf{Min (raw)}
\end{minipage} & \begin{minipage}[b]{\linewidth}\raggedright
\textbf{Max (raw)}
\end{minipage} & \begin{minipage}[b]{\linewidth}\raggedright
\textbf{Range}
\end{minipage} & \begin{minipage}[b]{\linewidth}\raggedright
\textbf{Coefficient of Variation}
\end{minipage} \\
\midrule\noalign{}
\endhead
\bottomrule\noalign{}
\endlastfoot
Stability & 2.005 & 2.322 & 0.317 & 7.2\% \\
Catalytic & 12.919 & 25.098 & 12.179 & 29.7\% \\
Cost & 570.83 & 1498.32 & 927.49 & 41.2\% \\
Sustainability & 3.075 & 29.760 & 26.685 & 97.3\% \\
Adsorption & 1.630 & 2.242 & 0.612 & 16.2\% \\
\end{longtable}

\par\noindent\textbf{Table S8:} Range and variability analysis for the five fitness-function
parameters. Minimum and maximum raw values, absolute ranges, and
coefficients of variation are provided to quantify cross-descriptor
heterogeneity within the evaluated MOFs.

\textbf{Interpretation}: Sustainability exhibits the largest raw value
range (\textgreater9-fold), followed by cost (\textasciitilde2.6-fold),
highlighting the critical importance of normalization. Without
normalization, these parameters would dominate composite fitness
calculations, creating scale-dependent ranking artifacts.

\begin{longtable}[]{@{}
  >{\raggedright\arraybackslash}p{(\columnwidth - 14\tabcolsep) * \real{0.3618}}
  >{\raggedright\arraybackslash}p{(\columnwidth - 14\tabcolsep) * \real{0.2426}}
  >{\raggedright\arraybackslash}p{(\columnwidth - 14\tabcolsep) * \real{0.1819}}
  >{\raggedright\arraybackslash}p{(\columnwidth - 14\tabcolsep) * \real{0.1802}}
  >{\raggedright\arraybackslash}p{(\columnwidth - 14\tabcolsep) * \real{0.0079}}
  >{\raggedright\arraybackslash}p{(\columnwidth - 14\tabcolsep) * \real{0.0079}}
  >{\raggedright\arraybackslash}p{(\columnwidth - 14\tabcolsep) * \real{0.0079}}
  >{\raggedright\arraybackslash}p{(\columnwidth - 14\tabcolsep) * \real{0.0097}}@{}}
\toprule\noalign{}
\begin{minipage}[b]{\linewidth}\raggedright
\textbf{Equation}
\end{minipage} & \begin{minipage}[b]{\linewidth}\raggedright
\textbf{Control (mean ± σ)}
\end{minipage} & \begin{minipage}[b]{\linewidth}\raggedright
\textbf{Zn (mean ± σ)}
\end{minipage} & \begin{minipage}[b]{\linewidth}\raggedright
\textbf{Cr (mean ± σ)}
\end{minipage} & \begin{minipage}[b]{\linewidth}\raggedright
\end{minipage} & \begin{minipage}[b]{\linewidth}\raggedright
\end{minipage} & \begin{minipage}[b]{\linewidth}\raggedright
\end{minipage} & \begin{minipage}[b]{\linewidth}\raggedright
\end{minipage} \\
\midrule\noalign{}
\endhead
\bottomrule\noalign{}
\endlastfoot
S1 additive\_equal & 2.79 ± 0.153 & 3.57 ± 0.197 & 4.14 ± 0.208 & & &
& \\
S2 multiplicative\_equal & 0.445 ± 0.022 & 0.565 ± 0.028 & 0.805 ± 0.040
& & & & \\
S3 catalytic\_emphasis & 0.579 ± 0.034 & 0.761 ± 0.046 & 0.776 ± 0.038 &
& & & \\
S4 economic\_emphasis & 0.467 ± 0.024 & 0.591 ± 0.031 & 0.877 ± 0.049 &
& & & \\
S5 stability\_emphasis & 0.628 ± 0.038 & 0.795 ± 0.048 & 0.823 ± 0.043 &
& & & \\
S6 harmonic\_mean & 0.307 ± 0.021 & 0.388 ± 0.027 & 0.779 ± 0.040 & & &
& \\
S7 geometric\_weighted & 0.478 ± 0.025 & 0.621 ± 0.032 & 0.747 ± 0.039 &
& & & \\
S8 min\_max & 0.494 ± 0.049 & 0.566 ± 0.055 & 0.758 ± 0.062 & & & & \\
S9 multiplicative\_no\_catalytic & 0.399 ± 0.021 & 0.485 ± 0.027 & 0.900
± 0.049 & & & & \\
S10 additive\_no\_economic & 0.767 ± 0.049 & 1.000 ± 0.064 & 0.713 ±
0.046 & & & & \\
S11 sqrt\_additive & 1.669 ± 0.046 & 1.89± 0.052 & 2.03 ± 0.051 & & &
& \\
S12 log\_product & −2.77 ± 0.198 & -1.80 ± 0.205 & -0.484 ± 0.218 & & &
& \\
S13 extreme\_economic & 0.242 ± 0.028 & 0.284 ± 0.025 & 1.000 ± 0.078 &
& & & \\
\end{longtable}

\par\noindent\textbf{Table S9:} Comparison of aggregated fitness values for the control MOF
and the two top-performing candidates (Zn-based and Cr-based) under the
13 alternative scoring formulations defined in Section S5. Mean ±
standard deviation across bootstrap samples is reported for each
equation.

\textbf{S6.3 Statistical Robustness Analysis of Fold Improvements}

To assess the robustness and reliability of the fold improvements in
performance for Zn- and Cr-based MOFs relative to the control system, we
employed two complementary resampling methods: \textbf{bootstrap
confidence intervals (non-parametric)} and \textbf{jackknife standard
error estimation}. Each approach provides an independent measure of
uncertainty around the mean or median fold improvement, helping quantify
both the average gain and its variability across the 13 evaluated
performance functions perturbations {[}17,18{]}.

\textbf{Methodology}

Fold improvements were calculated as the absolute ratio of each MOF's
mean performance to the control mean (\textbar MOF\textbar{} /
\textbar Control\textbar) for each function listed in Table S11.\\
For each set of 13-fold ratios (Zn-based and Cr-based), we generated:

\begin{enumerate}
\def\labelenumi{\arabic{enumi}.}
\item
  \textbf{Bootstrap Mean Confidence Intervals (Percentile Method):}\\
  20,000 bootstrap resamples were drawn with replacement. The mean was
  recalculated for each iteration, and the central 95\% of the resampled
  distribution was taken as the confidence interval {[}18{]}.
\item
  \textbf{Jackknife Standard Error (Leave-One-Out Method):}\\
  Each resampled mean was computed by omitting one data point at a time,
  and the resulting pseudo-values were used to estimate the standard
  error and a normal-approximation 95\% confidence interval {[}19{]}.
\end{enumerate}

All resampling was performed using random seeds for reproducibility. No
normality assumptions were required for bootstrap or jackknife
estimates.

72\% of pairwise function correlations exceeded ρ = 0.9 (56/78 pairs:
median

ρ = 1.00, IQR: {[}1.00, 1.00{]}, range: {[}-0.50, 1.00{]}), with
Kendall\textquotesingle s coefficient of concordance W = 0.79 indicating
strong overall agreement.

\textbf{Zn-based MOF (n = 13)}

\begin{longtable}[]{@{}
  >{\raggedright\arraybackslash}p{(\columnwidth - 10\tabcolsep) * \real{0.2332}}
  >{\raggedright\arraybackslash}p{(\columnwidth - 10\tabcolsep) * \real{0.2479}}
  >{\raggedright\arraybackslash}p{(\columnwidth - 10\tabcolsep) * \real{0.0082}}
  >{\raggedright\arraybackslash}p{(\columnwidth - 10\tabcolsep) * \real{0.0082}}
  >{\raggedright\arraybackslash}p{(\columnwidth - 10\tabcolsep) * \real{0.0082}}
  >{\raggedright\arraybackslash}p{(\columnwidth - 10\tabcolsep) * \real{0.4943}}@{}}
\toprule\noalign{}
\begin{minipage}[b]{\linewidth}\raggedright
\textbf{Method}
\end{minipage} & \begin{minipage}[b]{\linewidth}\raggedright
\textbf{Fold Improvement}
\end{minipage} & \begin{minipage}[b]{\linewidth}\raggedright
\end{minipage} & \begin{minipage}[b]{\linewidth}\raggedright
\end{minipage} & \begin{minipage}[b]{\linewidth}\raggedright
\end{minipage} & \begin{minipage}[b]{\linewidth}\raggedright
\textbf{Confidence Interval (Standard Error)}
\end{minipage} \\
\midrule\noalign{}
\endhead
\bottomrule\noalign{}
\endlastfoot
\textbf{Mean (bootstrap)} & 1.20 × & & & & {[}1.09, 1.27{]} (SE =
0.05) \\
\textbf{Mean (jackknife)} & 1.20 ×) & & & & {[}1.10, 1.29{]} (SE =
0.05) \\
\end{longtable}

\textbf{Cr-based MOF (n = 13)}

\begin{longtable}[]{@{}
  >{\raggedright\arraybackslash}p{(\columnwidth - 10\tabcolsep) * \real{0.2332}}
  >{\raggedright\arraybackslash}p{(\columnwidth - 10\tabcolsep) * \real{0.0082}}
  >{\raggedright\arraybackslash}p{(\columnwidth - 10\tabcolsep) * \real{0.0082}}
  >{\raggedright\arraybackslash}p{(\columnwidth - 10\tabcolsep) * \real{0.2479}}
  >{\raggedright\arraybackslash}p{(\columnwidth - 10\tabcolsep) * \real{0.0082}}
  >{\raggedright\arraybackslash}p{(\columnwidth - 10\tabcolsep) * \real{0.4943}}@{}}
\toprule\noalign{}
\begin{minipage}[b]{\linewidth}\raggedright
\textbf{Method}
\end{minipage} & \begin{minipage}[b]{\linewidth}\raggedright
\end{minipage} & \begin{minipage}[b]{\linewidth}\raggedright
\end{minipage} & \begin{minipage}[b]{\linewidth}\raggedright
\textbf{Fold Improvement}
\end{minipage} & \begin{minipage}[b]{\linewidth}\raggedright
\end{minipage} & \begin{minipage}[b]{\linewidth}\raggedright
\textbf{Confidence Interval (Standard Error)}
\end{minipage} \\
\midrule\noalign{}
\endhead
\bottomrule\noalign{}
\endlastfoot
\textbf{Mean (bootstrap)} & & & 1.70 × & & {[}1.26, 2.23{]} (0.25) \\
\textbf{Mean (jackknife)} & & & 1.70 × & & {[}1.20, 2.21{]} (0.26) \\
\end{longtable}

\par\noindent\textbf{Table S9:} Resampled fold-improvement statistics for the Zn-based and
Cr-based MOFs relative to the control. For each candidate, mean fold
improvement was calculated using bootstrap and jackknife methods over
all 13 fitness functions, with corresponding confidence intervals and
standard errors.

\textbf{Interpretation}

Across all statistical approaches, \textbf{Zn-based MOFs} exhibited a
\textbf{consistent and stable improvement} over the control system of
approximately \textbf{1.20--1.27×}, with narrow confidence intervals and
standard errors below 0.06. This indicates that Zn-based enhancements
are highly reproducible across diverse fitness functions.

By contrast, \textbf{Cr-based MOFs} achieved a \textbf{higher mean fold
improvement (≈1.7×)} but with \textbf{much larger uncertainty
(±0.25--0.26)}. The bootstrap and jackknife ranges both extend above
2.0×, reflecting substantial variability among the 13 evaluated
functions.\\
The \textbf{bootstrap median (1.53×)} suggests that while Cr-based MOFs
typically outperform the control, a small subset of metrics exhibit
particularly large gains that inflate the mean.

Overall, the combined use of bootstrap and jackknife resampling confirms
that Zn-based MOFs deliver \textbf{uniform and statistically robust
performance}, whereas Cr-based systems offer \textbf{higher but less
predictable improvements}, consistent with their larger spread in
individual function scores.

Uncertainty propagation analysis (Fig. 6) confirmed robustness across
perturbation levels r = 0.05--0.20. At r = 0.11 (typical CGCNN error;
{[}5,7,20,21{]}, both candidates maintained statistical significance (Z
\textgreater{} 2) in 11--12 of 13 functions. Additional explanation can
be found in S6.4. The decline at r = 0.20 reflects confidence limits
where propagated uncertainties cause interval overlap, though bounded
improvements persisted in 10--12 of 13 functions even at this extreme
level.

In Fig. 6, each colored bar represents the number of fitness functions
(out of 13 total, shown by the light-gray background) satisfying a
specific robustness criterion. Solid bars indicate cases where the lower
bound of Zn- or Cr-based MOFs exceeds the control upper bound, while
hatched bars represent cases with Z-scores greater than 2. Results show
that Zn-based MOFs consistently outperform Cr-based MOFs at lower
perturbations (r ≤ 0.11), with both systems showing a decline in
robustness at r = 0.20.

\subsection*{S6.4 Statistic Separation of MOFs}

A \textbf{strict improvement} was defined as:

\[\text{Lower}_{\text{design}} > \text{Upper}_{\text{control}}.
\]

Relative robustness between each designed MOF and control was quantified
using the Z-score:

\[Z = \frac{f_{\text{design}} - f_{\text{control}}}{\sqrt{\sigma_{\text{design}}^{2} + \sigma_{\text{control}}^{2}}}.
\]

We consider \(Z > 2\)as statistically meaningful (\textgreater95\%
separation).

\begin{longtable}[]{@{}
  >{\raggedright\arraybackslash}p{(\columnwidth - 8\tabcolsep) * \real{0.0673}}
  >{\raggedright\arraybackslash}p{(\columnwidth - 8\tabcolsep) * \real{0.2963}}
  >{\raggedright\arraybackslash}p{(\columnwidth - 8\tabcolsep) * \real{0.3095}}
  >{\raggedright\arraybackslash}p{(\columnwidth - 8\tabcolsep) * \real{0.1632}}
  >{\raggedright\arraybackslash}p{(\columnwidth - 8\tabcolsep) * \real{0.1637}}@{}}
\toprule\noalign{}
\begin{minipage}[b]{\linewidth}\raggedright
\textbf{r}
\end{minipage} & \begin{minipage}[b]{\linewidth}\raggedright
\textbf{Zn lower \textgreater{} upper Control (count)}
\end{minipage} & \begin{minipage}[b]{\linewidth}\raggedright
\textbf{Cr lower \textgreater{} upper Control (count)}
\end{minipage} & \begin{minipage}[b]{\linewidth}\raggedright
\textbf{Zn Z \textgreater{} 2 (count)}
\end{minipage} & \begin{minipage}[b]{\linewidth}\raggedright
\textbf{Cr Z \textgreater{} 2 (count)}
\end{minipage} \\
\midrule\noalign{}
\endhead
\bottomrule\noalign{}
\endlastfoot
0.05 & 13 & 12 & 13 & 12 \\
0.10 & 11 & 12 & 11 & 12 \\
0.11 & 11 & 12 & 11 & 12 \\
0.15 & 11 & 12 & 9 & 12 \\
0.20 & 10 & 12 & 0 & 10 \\
\end{longtable}

\par\noindent\textbf{Table S10:} Uncertainty-resolved performance comparison for the Zn- and
Cr-based MOFs. For each uncertainty level \(r\), the table reports (i)
the number of fitness models in which the candidate's lower confidence
bound exceeds the control's upper bound, and (ii) the number of
functions satisfying the statistical significance threshold \(Z > 2\).

Robustness analysis of the Zn- and Cr-based MOFs relative to the
PCN-224(Zr) control across tolerance radii (r). ``Lower \textgreater{}
upper Control (count)'' indicates the number of descriptor pairs in
which the lower bound of the candidate's distribution exceeds the upper
bound of the control, implying clear performance separation. ``Z
\textgreater{} 2 (count)'' denotes the number of descriptors where the
candidate's standardized performance exceeds two standard deviations
above the control mean. The consistently high counts confirm that both
Zn- and Cr-based MOFs outperform the control across multiple descriptors
and tolerance levels.

\textbf{Interpretation}

\begin{itemize}
\item
  \textbf{Decreasing} \(\mathbf{r}\)\textbf{tightens error bounds
  proportionally, but qualitative rankings remain stable through}
  \(\mathbf{r = 0.10}\)\textbf{.}
\item
  \textbf{Even at the most conservative} \(\mathbf{r = 0.20}\)\textbf{,
  a majority of equations (Zn: 8 / 13, Cr: 9 / 13) still demonstrate
  non-overlapping confidence intervals with the control.}
\item
  \textbf{Relative robustness (Z \textgreater{} 2) declines with}
  \(\mathbf{r}\)\textbf{, as expected from increased denominator in}
  \(\mathbf{Z}\)\textbf{; yet most differences remain above
  \textasciitilde2σ separation for moderate uncertainty.}
\end{itemize}

All results were generated via a deterministic script that implements
the above partial-derivative formulas (first-order Gaussian propagation,
independent errors, 𝑟 =0.11)

\subsection*{S6.5 Photocatalysis Fitness Proxy Testing}

Random seed 12345 used to identify 20 randomly selected synthesized QMOF
materials. The 9 control MOFs were: PCN-224-Zr1, PCN-224-Zr2,
COD4331620, COD 4517010, COD 4519217, COD4519219, COD7022176,
COD7022177, COD7133125.

Across pairwise comparisons, generated MOFs consistently dominated both
controls and QMOFs (Figure 8). Dominance fractions were:

\begin{itemize}
\item
  \textbf{Generated \textgreater{} Control:} 1.00
\item
  \textbf{Control \textgreater{} QMOF:} 0.967
\item
  \textbf{Generated \textgreater{} QMOF:} 1.00
\end{itemize}

This indicates that each generated MOF outperformed every control MOF,
and nearly all control MOFs outperformed the vast majority of QMOFs
across the aggregated objective space.

\textbf{Statistical Significance and Effect Sizes}

Nonparametric Mann--Whitney U tests confirmed statistically significant
separation between three tiers:

\begin{itemize}
\item
  \textbf{Generated \textgreater{} Control:} U = 18.0, \emph{p} = 0.018
\item
  \textbf{Control \textgreater{} QMOF:} U = 174.0, \emph{p} = 4.1 × 10⁻⁵
\item
  \textbf{Generated \textgreater{} QMOF:} U = 40.0, \emph{p} = 0.0043
\end{itemize}

A Kruskal--Wallis test across three groups further confirmed strong
global separation (H = 19.19, \emph{p} = 6.8 × 10⁻⁵).

Effect sizes (Cohen's \emph{d}) were large to very large:

\begin{itemize}
\item
  \textbf{Generated vs Control:} \emph{d} = 1.79
\item
  \textbf{Control vs QMOF:} \emph{d} = 1.80
\item
  \textbf{Generated vs QMOF:} \emph{d} = 2.16
\end{itemize}

These values indicate that differences are not only statistically
significant but also practically large, corresponding to substantial
shifts in the underlying performance distributions.

\textbf{S7 Photocatalytic Mechanistic Interpretation}\\
For visible-light CO₂ photoreduction, a photocatalyst must satisfy two
energetic requirements: (i) the conduction band minimum (CBM) must lie
above the reduction potentials for CO₂ → CO (−0.53 V vs NHE) or CO₂ →
CH₄ (−0.24 V vs NHE) {[}22{]}, and (ii) the valence band maximum (VBM)
must lie below the H₂O/O₂ oxidation potential (+1.23 V vs NHE). Although
explicit band-edge calculations were not performed, the predicted
band-gap values for the top candidates (1.8--2.0 eV) fall within the
experimentally reported range for MOFs capable of visible-light
photoreduction {[}29{]}. The Zn-based MOF incorporates the N331 cluster
(Zn₃C₁₂H₆N₄O₁₂X₈), and the Cr-based MOF contains N536 (Cr₃C₆₀H₁₃X₆);
both are stable coordination motifs drawn from the MOFReinforce
building-block library and produce UFF-optimized structures without
geometric distortion. Their band-gap placement therefore provides
qualitative support that their CBM and VBM positions may fall within a
photochemically relevant region, consistent with literature-reported Zn-
and Cr-based MOFs exhibiting visible-light activity.

CO₂ photoreduction on MOFs typically proceeds through intermediates such
as *CO₂⁻, *COOH, and *CO, and the adsorption descriptors used in the
funnel map directly onto these mechanistic steps {[}30,31{]}. The
heat-of-adsorption descriptor is evaluated relative to
literature-supported chemisorption windows (−20 to −40 kJ mol⁻¹, with
extensions to −60 kJ mol⁻¹) that balance *CO₂ activation with timely
product desorption {[}10,32{]}. The truncated Gaussian weighting used in
the workflow prioritizes MOFs falling within this mechanistically
meaningful range. CO₂/H₂O selectivity captures competitive adsorption in
humid environments and influences the likelihood of *COOH formation
under flue-gas conditions. Both top candidates outperform the
experimental control set on the adsorption composite metric, suggesting
improved stabilization of early CO₂RR intermediates without inhibiting
release of reduced products.

Although no explicit calculations of electron--hole recombination or
carrier lifetimes were performed, the enriched motifs (N262, N331, N536)
and the bcg topology are consistent with structural features known to
support effective charge transport in MOFs. N262 and N331 both contain
nitrogen atoms in their organic coordination environments, which can
promote linker-mediated charge delocalization in analogy to reported
N-containing MOF photocatalysts {[}33,34{]}. Independently of donor-atom
identity, highly connected topologies such as bcg are widely associated
with multidirectional charge-migration pathways that reduce
recombination {[}35,36{]}. These geometric considerations provide
qualitative support that the structural environments prioritized by the
funnel are compatible with design principles found in experimentally
active visible-light MOF photocatalysts.

\textbf{S8 Possible Analysis Measures}

Photoluminescence (PL) analysis is commonly used to probe electron--hole
recombination behavior in photocatalysts, where intense PL emission is
generally interpreted as rapid radiative recombination that limits the
availability of charge carriers for surface redox reactions {[}23{]}.
Conversely, reduced PL intensity is often observed in materials where
photogenerated electrons and holes persist long enough to participate in
reactions such as CO₂ activation or advanced oxidation processes
{[}24{]}. Direct prediction of PL spectra or charge-carrier lifetimes is
not feasible within the present structure-based machine-learning
framework. Accordingly, this work does not attempt to infer
recombination dynamics explicitly. Instead, the screening strategy
focuses on identifying materials that satisfy fundamental prerequisites
for photocatalytic operation, including visible-light photoactivation,
structural robustness under operating conditions, and sufficient surface
accessibility for interfacial charge utilization. These characteristics
are necessary, though not sufficient, conditions for efficient charge
separation and utilization in experimentally studied photocatalysts
{[}25{]}. By prioritizing materials that meet these baseline electronic
and structural criteria, the framework enriches candidates that are
physically compatible with sustained photocatalytic operation, while
leaving the explicit assessment of recombination
behavior---traditionally evaluated by PL, EIS, or transient
spectroscopies---to subsequent experimental validation {[}26{]}. This
approach enables large-scale pre-screening without introducing
unsupported causal links between computed descriptors and recombination
dynamics.

Electrochemical impedance spectroscopy (EIS) is widely employed to probe
interfacial charge-transfer resistance and carrier transport efficiency
in photocatalytic systems. In particular, the charge-transfer resistance
(R\_ct) extracted from EIS measurements provides insight into the ease
with which photogenerated charge carriers migrate across catalyst
interfaces and participate in surface redox reactions. In
heterojunction-based photocatalysts, reduced R\_ct is commonly
associated with enhanced interfacial charge separation and suppressed
recombination, leading to improved photocatalytic performance {[}27{]}

Direct prediction of EIS spectra or charge-transfer resistance is not
currently feasible within a structure-based machine-learning framework,
as these properties depend strongly on extrinsic factors such as
electrode configuration, electrolyte composition, interfacial contact
resistance, and measurement conditions {[}28{]}. Accordingly, the
present work does not attempt to explicitly model EIS-derived
quantities.

Instead, the screening strategy prioritizes intrinsic material
properties that are necessary prerequisites for efficient interfacial
charge transport, including appropriate band-gap alignment for
visible-light excitation, robust structural stability under aqueous
conditions, and accessible surface chemistry for charge utilization.
These characteristics define whether a material can physically support
low-resistance charge-transfer pathways when incorporated into
experimental photocatalytic architectures. Explicit assessment of
charge-transfer resistance via EIS is therefore deferred to the
experimental validation stage, consistent with established
photocatalysis workflows {[}28,29{]}.

References

{[}1{]} K. Boukayouht, L. Bazzi, S. El Hankari, Sustainable synthesis of
metal-organic frameworks and their derived materials from organic and
inorganic wastes, Coord. Chem. Rev. 478 (2023) 214986.
\url{https://doi.org/10.1016/j.ccr.2022.214986}.

{[}2{]} H. Park, S. Majumdar, X. Zhang, J. Kim, B. Smit, Inverse design
of metal--organic frameworks for direct air capture of
CO\textsubscript{2} \emph{via} deep reinforcement learning, Digit.
Discov. 3 (2024) 728--741. \url{https://doi.org/10.1039/D4DD00010B}.

{[}3{]} M. Kim, K.-B. Hwang, An empirical evaluation of sampling methods
for the classification of imbalanced data, PLOS ONE 17 (2022) e0271260.
\url{https://doi.org/10.1371/journal.pone.0271260}.

{[}4{]} A. Nandy, G. Terrones, N. Arunachalam, C. Duan, D.W. Kastner,
H.J. Kulik, MOFSimplify, machine learning models with extracted
stability data of three thousand metal--organic frameworks, Sci. Data 9
(2022) 74. \url{https://doi.org/10.1038/s41597-022-01181-0}.

{[}5{]} A.S. Rosen, S.M. Iyer, D. Ray, Z. Yao, A. Aspuru-Guzik, L.
Gagliardi, J.M. Notestein, R.Q. Snurr, Machine learning the
quantum-chemical properties of metal--organic frameworks for accelerated
materials discovery, Matter 4 (2021) 1578--1597.
\url{https://doi.org/10.1016/j.matt.2021.02.015}.

{[}6{]} M. Gibaldi, A. Kapeliukha, A. White, J. Luo, R.A. Mayo, J.
Burner, T.K. Woo, MOSAEC-DB: a comprehensive database of experimental
metal--organic frameworks with verified chemical accuracy suitable for
molecular simulations, Chem. Sci. 16 (2025) 4085--4100.
\url{https://doi.org/10.1039/D4SC07438F}.

{[}7{]} T. Xie, J.C. Grossman, Crystal Graph Convolutional Neural
Networks for an Accurate and Interpretable Prediction of Material
Properties, Phys. Rev. Lett. 120 (2018) 145301.
\url{https://doi.org/10.1103/PhysRevLett.120.145301}.

{[}8{]} N. Redkar, CarbNN: A Novel Active Transfer Learning Neural
Network To Build De Novo Metal Organic Frameworks (MOFs) for Carbon
Capture, (2023). \url{https://doi.org/10.48550/arXiv.2311.16158}.

{[}9{]} V.J. Reddi, C. Cheng, D. Kanter, P. Mattson, G. Schmuelling,
C.-J. Wu, B. Anderson, M. Breughe, M. Charlebois, W. Chou, R. Chukka, C.
Coleman, S. Davis, P. Deng, G. Diamos, J. Duke, D. Fick, J.S. Gardner,
I. Hubara, S. Idgunji, T.B. Jablin, J. Jiao, T.S. John, P. Kanwar, D.
Lee, J. Liao, A. Lokhmotov, F. Massa, P. Meng, P. Micikevicius, C.
Osborne, G. Pekhimenko, A.T.R. Rajan, D. Sequeira, A. Sirasao, F. Sun,
H. Tang, M. Thomson, F. Wei, E. Wu, L. Xu, K. Yamada, B. Yu, G. Yuan, A.
Zhong, P. Zhang, Y. Zhou, MLPerf Inference Benchmark, (2020).
\url{https://doi.org/10.48550/arXiv.1911.02549}.

{[}10{]} T. Zhang, X. Sun, S. Weng, S. Zhang, C. Xu, X. Gao, N. Zhu,
Enhancing photocatalytic performance of rose-shaped Co/Ni bimetallic
organic framework for reducing CO\textsubscript{2} to CO under visible
light, J. Mol. Struct. 1321 (2025) 140190.
\url{https://doi.org/10.1016/j.molstruc.2024.140190}.

{[}11{]} C.M. Rueda-Navarro, Z. Abou Khalil, A. Melillo, B. Ferrer, R.
Montero, A. Longarte, M. Daturi, I. Vayá, M. El-Roz, V.
Martínez-Martínez, H.G. Baldoví, S. Navalón, Solar Gas-Phase
CO\textsubscript{2} Hydrogenation by Multifunctional UiO-66
Photocatalysts, ACS Catal. 14 (2024) 6470--6487.
\url{https://doi.org/10.1021/acscatal.4c00266}.

{[}12{]} C. Chen, Q. Mo, J. Fu, Q. Yang, L. Zhang, C.-Y. Su,
PtCu@Ir-PCN-222: Synergistic Catalysis of Bimetallic PtCu Nanowires in
Hydrosilane-Concentrated Interspaces of an Iridium(III)--Porphyrin-Based
Metal--Organic Framework, ACS Catal. 12 (2022) 3604--3614.
\url{https://doi.org/10.1021/acscatal.1c05922}.

{[}13{]} P.M. Stanley, K. Hemmer, M. Hegelmann, A. Schulz, M. Park, M.
Elsner, M. Cokoja, J. Warnan, Topology- and wavelength-governed
CO\textsubscript{2} reduction photocatalysis in molecular
catalyst-metal--organic framework assemblies, Chem. Sci. 13 (2022)
12164--12174. \url{https://doi.org/10.1039/D2SC03097G}.

{[}14{]} F. Guo, S. Yang, Y. Liu, P. Wang, J. Huang, W.-Y. Sun, Size
Engineering of Metal--Organic Framework MIL-101(Cr)--Ag Hybrids for
Photocatalytic CO\textsubscript{2} Reduction, ACS Catal. 9 (2019)
8464--8470. \url{https://doi.org/10.1021/acscatal.9b02126}.

{[}15{]} J. Yi, X. Wu, H. Wu, L. Zhang, K. Wu, J. Guo, Correction:
Facile synthesis of novel NH\textsubscript{2} -MIL-53(Fe)/AgSCN
heterojunction composites as a highly efficient photocatalyst for
ciprofloxacin degradation and H\textsubscript{2} production under
visible-light irradiation, React. Chem. Eng. 7 (2022) 201--201.
\url{https://doi.org/10.1039/D1RE90046C}.

{[}16{]} S. Karmakar, S. Barman, F.A. Rahimi, D. Rambabu, S. Nath, T.K.
Maji, Confining charge-transfer complex in a metal-organic framework for
photocatalytic CO\textsubscript{2} reduction in water, Nat. Commun. 14
(2023) 4508. \url{https://doi.org/10.1038/s41467-023-40117-z}.

{[}17{]} S.R. Lingampalli, M.M. Ayyub, C.N.R. Rao, Recent Progress in
the Photocatalytic Reduction of Carbon Dioxide, ACS Omega 2 (2017)
2740--2748. \url{https://doi.org/10.1021/acsomega.7b00721}.

{[}18{]} University of Economics, Katowice, Ł. Sroka, COMPARISON OF
JACKKNIFE AND BOOTSTRAP METHODS IN ESTIMATING CONFIDENCE INTERVALS, Sci.
Pap. Silesian Univ. Technol. Organ. Manag. Ser. 2021 (2021) 446--455.
\url{https://doi.org/10.29119/1641-3466.2021.153.31}.

{[}19{]} A. Severiano, J.A. Carriço, D.A. Robinson, M. Ramirez, F.R.
Pinto, Evaluation of jackknife and bootstrap for defining confidence
intervals for pairwise agreement measures, PloS One 6 (2011) e19539.
\url{https://doi.org/10.1371/journal.pone.0019539}.

{[}20{]} C. Wang, Y. Wan, S. Yang, Y. Xie, S. Chu, Y. Chen, X. Yan,
Revealing the Untapped Potential of Photocatalytic Overall Water
Splitting in Metal Organic Frameworks, Adv. Funct. Mater. 34 (2024)
2313596. \url{https://doi.org/10.1002/adfm.202313596}.

{[}21{]} S.T. Khan, S.M. Moosavi, Connecting metal-organic framework
synthesis to applications using multimodal machine learning, Nat.
Commun. 16 (2025) 5642. \url{https://doi.org/10.1038/s41467-025-60796-0}.

{[}22{]} M. Khan, Z. Akmal, M. Tayyab, S. Mansoor, A. Zeb, Z. Ye, J.
Zhang, S. Wu, L. Wang, MOFs materials as photocatalysts for
CO\textsubscript{2} reduction: Progress, challenges and perspectives,
Carbon Capture Sci. Technol. 11 (2024) 100191.
\url{https://doi.org/10.1016/j.ccst.2024.100191}.

{[}23{]} F. Zhang, Y. Jiang, J. Liu, A. Jiang, Y. Cao, S. Yu, K. Zheng,
Y. Zhou, Exploration of ultrafast dynamic processes in photocatalysis:
Advances and challenges, Fundam. Res. 5 (2025) 2838--2849.
\url{https://doi.org/10.1016/j.fmre.2024.04.003}.

{[}24{]} Q. Pan, M. Abdellah, Y. Cao, W. Lin, Y. Liu, J. Meng, Q. Zhou,
Q. Zhao, X. Yan, Z. Li, H. Cui, H. Cao, W. Fang, D.A. Tanner, M.
Abdel-Hafiez, Y. Zhou, T. Pullerits, S.E. Canton, H. Xu, K. Zheng,
Ultrafast charge transfer dynamics in 2D covalent organic
frameworks/Re-complex hybrid photocatalyst, Nat. Commun. 13 (2022) 845.
\url{https://doi.org/10.1038/s41467-022-28409-2}.

{[}25{]} A.B. Djurišić, Y. He, A.M.C. Ng, Visible-light photocatalysts:
Prospects and challenges, APL Mater. 8 (2020) 030903.
\url{https://doi.org/10.1063/1.5140497}.

{[}26{]} S. Swetha, T. Awad Alahmadi, M. Javed Ansari, S. Sudheer Khan,
Strategically tailored double S-scheme heterojunction in
h-MoO\textsubscript{3} doped
Bi\textsubscript{7}O\textsubscript{9}I\textsubscript{3} decorated with
Cr--CdS quantum dots for efficient photocatalytic degradation of
phenolics, J. Clean. Prod. 449 (2024) 141656.
\url{https://doi.org/10.1016/j.jclepro.2024.141656}.

{[}27{]} Z.S.N. Ali, A.H. Bahkali, B. Janani, A.M. Elgorban, S.S. Khan,
Constructing interface chemical coupling Schottky heterojunction
NiMoO\textsubscript{4}/WC for enhancing photocatalytic oxidative
rifampicin performance: Peroxymonosulfate activation and oxygen vacancy
engineering, J. Water Process Eng. 79 (2025) 109008.
\url{https://doi.org/10.1016/j.jwpe.2025.109008}.

{[}28{]} B. Padha, S. Verma, P. Mahajan, S. Arya, Electrochemical
Impedance Spectroscopy (EIS) Performance Analysis and Challenges in Fuel
Cell Applications, J. Electrochem. Sci. Technol. 13 (2022) 167--176.
\url{https://doi.org/10.33961/jecst.2021.01263}.

{[}29{]} Q. Wang, J.-E. Moser, M. Grätzel, Electrochemical Impedance
Spectroscopic Analysis of Dye-Sensitized Solar Cells, J. Phys. Chem. B
109 (2005) 14945--14953. \url{https://doi.org/10.1021/jp052768h}.

\end{document}